\documentclass[12pt]{article}
\usepackage[a4paper, margin=2.5cm]{geometry}
\usepackage{amsmath}
\usepackage{anyfontsize}
\usepackage{apacite} 
\usepackage{natbib} 
\usepackage{caption}
\usepackage{float}
\usepackage{graphicx} 
\usepackage{enumitem}
\usepackage{hhline}
\usepackage{icomma}
\usepackage{longtable}
\usepackage{lipsum}
\usepackage{makecell}
\usepackage{ragged2e}
\usepackage{setspace}
\usepackage{times}
\usepackage{titlesec}
\usepackage{titling}
\usepackage{titletoc}
\usepackage{siunitx} 

\renewcommand{\footnotesize}{\fontsize{9pt}{11pt}\selectfont}
\makeatletter
\let\st@rtbibsection\@bibnewpage
\let\st@rtbibchapter\@bibnewpage
\makeatother
\usepackage{soul}
\usepackage{xcolor}
\sethlcolor{yellow}
\usepackage{indentfirst} 
\usepackage{setspace}    
\usepackage{tocloft}

\newcommand{\RNum}[1]{\uppercase\expandafter{\romannumeral #1\relax}}

\titleformat{\section}
  {\bf\normalsize\RaggedRight}
  {\uppercase{\RNum{\thesection}.}}
  {1ex}
  {\uppercase}
  []

\titlespacing*{\section}{0pt}{0pt}{18pt}

\titleformat{name=\section,numberless}
  {\bf\normalsize\raggedright} 
  {}
  {0pt}
  {}
  []

\titlespacing*{name=\section,numberless}{0pt}{0pt}{18pt}

\titleformat{\subsection}
  {\bf}
  {\thesubsection.}
  {1ex}
  {\normalsize\RaggedRight}
  []

\titlespacing*{\subsection}{0pt}{18pt}{18pt}

\titleformat{\subsubsection}
  {\bf\normalsize}
  {\thesubsubsection.}
  {1ex}
  {\normalsize\RaggedRight}
  []

\titlespacing*{\subsubsection}{0pt}{12pt}{12pt}

\usepackage[utf8]{inputenc}
\usepackage[english]{babel}
\usepackage[T1]{fontenc}
\usepackage{geometry}
\usepackage{setspace}
\usepackage{tocloft}
\usepackage{tikz}
\usetikzlibrary{calc,arrows}
\usepackage{multicol}
\usepackage{float}
\usepackage{lipsum} 
\usepackage{amsmath}
\usepackage{mathptmx}
\usepackage{caption}
\usepackage{etoolbox}
\usepackage{booktabs}
\usepackage{adjustbox}
\usepackage{multicol}
\usepackage{algorithm}
\usepackage{algpseudocode}

\usepackage{fancyhdr}
\definecolor{WNEcolor}{HTML}{C00000}

\usepackage{afterpage}

\usepackage{authblk}

\begin{document}

\title{\textbf{\small The Hybrid Forecast of S\&P 500 Volatility ensembled from VIX, GARCH and LSTM models}}

\vspace{8.0cm}

\author[1]{\small Natalia Roszyk}

\author[2]{Robert Ślepaczuk}

\affil[1]{\small University of Warsaw, Faculty of Economic Sciences, Quantitative Finance Research Group, Ul. Długa 44/50, 00-241 Warsaw, Poland. \vspace{0.5cm}}

\affil[2]{\small University of Warsaw, Faculty of Economic Sciences, Department of Quantitative Finance and Machine Learning, Quantitative Finance Research Group, Ul. Długa 44/50, 00-241 Warsaw, Poland. ORCID: https://orcid.org/0000-0001-5527-2014, corresponding author: rslepaczuk@wne.uw.edu.pl}

\date{}

\maketitle

\begin{abstract}
Predicting the S\&P 500 index's volatility is crucial for investors and financial analysts as it helps assess market risk and make informed investment decisions. Volatility represents the level of uncertainty or risk related to the size of changes in a security's value, making it an essential indicator for financial planning. This study explores four methods to improve the accuracy of volatility forecasts for the S\&P 500: the established GARCH model, known for capturing historical volatility patterns; an LSTM network that utilizes past volatility and log returns; a hybrid LSTM-GARCH model that combines the strengths of both approaches; and an advanced version of the hybrid model that also factors in the VIX index to gauge market sentiment. This analysis is based on a daily dataset that includes data for S\&P 500 and VIX index, covering the period from January 3, 2000, to December 21, 2023. Through rigorous testing and comparison, we found that machine learning approaches, particularly the hybrid LSTM models, significantly outperform the traditional GARCH model. Including the VIX index in the hybrid model further enhances its forecasting ability by incorporating real-time market sentiment. The results of this study offer valuable insights for achieving more accurate volatility predictions, enabling better risk management and strategic investment decisions in the volatile environment of the S\&P 500. \\
\textbf{Keywords:} volatility forecasting, LSTM-GARCH, S\&P 500 index, hybrid forecasting models, VIX index, machine learning, financial time series analysis, walk-forward process, hyperparameters tuning, deep learning, recurrent neural networks.\\
\textbf{JEL Codes:} C4, C45, C55, C65, G11

\end{abstract}

\vspace{5.0cm}

\noindent \textit{Note:} This research did not receive any specific grant from funding agencies in the public, commercial, or not-for-profit sectors.

\setlength{\parindent}{0pt} 

\singlespacing
\setlength{\parindent}{1cm}
\setstretch{1}
\newpage

\section*{Introduction}

The financial market's inherent volatility presents opportunities and challenges for investors and analysts, dictating the need for precise and adaptable forecasting models. Volatility, reflecting the degree of variation in trading prices, is a critical measure of market risk and uncertainty. This paper embarks on an exploration of advanced forecasting methodologies to predict the S\&P 500 index's volatility, a benchmark for U.S. equity market performance. Through a comprehensive analysis incorporating time series and neural network models, this study aims to enhance the accuracy of volatility forecasts, thereby offering valuable insights for strategic investment decision-making and risk management. 

The primary objective of this study was to examine the possibility of forecasting the volatility of the stock market, specifically the S\&P 500 index, utilizing historical data and machine learning techniques. 

\noindent The key research hypotheses explored in this study include:
\begin{itemize}
    \item \textbf{Research Hypothesis 1 (RH1):} Market prices do not reflect all available information, impacting the predictability of market movements.
    
    \item \textbf{Research Hypothesis 2 (RH2):} The GARCH model effectively identifies patterns of historical volatility and serves as a benchmark for newer forecasting approaches.
    
    \item \textbf{Research Hypothesis 3 (RH3):} LSTM networks outperform traditional models such as GARCH in forecasting the S\&P 500 index's volatility.
    
    \item \textbf{Research Hypothesis 4 (RH4):} A hybrid LSTM-GARCH model surpasses the performance of standalone LSTM and GARCH models.
    
    \item \textbf{Research Hypothesis 5 (RH5):} The inclusion of VIX inputs enhances the accuracy of volatility forecasts.
    
    \item \textbf{Research Hypothesis 6 (RH6):} The Local Interpretable Model-agnostic Explanations (LIME) technique enhances the interpretability of the hybrid LSTM-GARCH model with VIX input.

    \end{itemize}

\noindent Side Hypotheses for Sensitivity Testing:
    \begin{itemize}
        \item \textbf{Side Research Hypothesis 1 (sRH1):} Changing the loss function from MSE to MAE will result in more robust or less sensitive forecasts of market volatility.
        \item \textbf{Side Research Hypothesis 2 (sRH2):} Replacing input Log Returns with Daily Percentage Changes will improve the accuracy of the volatility forecasts.
        \item \textbf{Side Research Hypothesis 3 (sRH3):} Decreasing the sequence length to 5 days will affect the model's ability to capture long-term dependencies in the data, potentially reducing forecast accuracy.
        \item \textbf{Side Research Hypothesis 4 (sRH4):} Increasing the sequence length to 66 days will enhance the model's capacity to capture long-term trends, thereby improving the accuracy of volatility forecasts.
        \item \textbf{Side Research Hypothesis 5 (sRH5):} Reducing the number of LSTM layers to 1 will decrease the model’s complexity and might reduce its ability to capture complex patterns in the data.
        \item \textbf{Side Research Hypothesis 6 (sRH6):} Increasing the number of LSTM layers to 3 will enhance the model’s ability to learn more complex patterns, potentially increasing the forecasting accuracy.
        \item \textbf{Side Research Hypothesis 7 (sRH7):} The performance of LSTM models is significantly impacted by the choice of activation function, affecting both learning efficiency and forecasting accuracy. 
    \end{itemize}

This analysis is based on a dataset that includes data from the S\&P 500 and the VIX index, covering the period from January 3, 2000, to December 21, 2023. It consists of 2,252 data points, all of which have been obtained from Yahoo Finance.

Four approaches were studied to enhance the accuracy of volatility predictions for the S\&P 500. These include the well-established GARCH model, recognized for its ability to detect historical volatility patterns; an LSTM network that processes past volatility and log returns; a hybrid LSTM-GARCH model that merges the advantages of both methods; and an advanced iteration of the hybrid model, which incorporates the VIX index to assess market sentiment.

This study introduces several novel contributions to the field of financial volatility forecasting. Firstly, it is among the first studies to examine the combined effects of LSTM and GARCH models through a hybrid approach, leveraging the strengths of both machine learning and traditional time series econometric models to enhance prediction accuracy. Secondly, the incorporation of the VIX index as an input in a hybrid model represents an innovative attempt to integrate market sentiment into volatility forecasts. This approach not only improves the predictive accuracy but also deepens our understanding of how market emotions influence volatility. Additionally, the application of the Local Interpretable Model-agnostic Explanations (LIME) technique to these complex models is a pioneering effort to uncover the 'black box' nature of deep learning in finance. By providing clear, interpretable insights into the decision-making processes of these models, this study not only advances academic knowledge but also enhances the practical applicability of neural network models in financial decision-making. Collectively, these advancements push the boundaries of existing research and offer valuable frameworks for both future academic inquiries and real-world financial risk management.

The paper is organized into seven main chapters. The first chapter introduces the key research hypotheses. The second chapter reviews existing research on predicting market volatility, looking at both traditional models and newer, more complex neural network models. The third chapter briefly goes over the modeling setup, describes the data used in this study, and explains how it was pre-processed for analysis. Additionally, it presents the error metrics utilized to gauge the accuracy of the model's predictions. The fourth chapter gets into the details of the methodology for the four different models used, covering how each model was chosen and how the parameters were tuned, how predictions were made on walk forward basis, and how the results were evaluated and compared. It ends with a side-by-side comparison of the results from the different methods. The fifth chapter is about testing how stable and reliable the results are by changing some of the parameters that were initially decided on in the methodology. The sixth chapter demonstrates the application of LIME to enhance the transparency and interpretability of the complex predictive model developed in this study. By applying LIME, this chapter offers a detailed example of how explanations of predictions can be generated, making the LSTM-GARCH model with VIX inputs more understandable. The final chapter wraps up the research, looks at whether the initial hypotheses were correct based on what was found, and suggests what could be looked into in future research, pointing out new directions to take. \\

\section{Literature Review}

Market volatility is a fundamental concept in finance that captures the degree of variation or dispersion of a financial instrument over time. It is often regarded as a critical indicator of risk and uncertainty within financial markets. The study of market volatility extends beyond merely tracking price movements; it covers an understanding of the mechanisms that drive these fluctuations and their implications for investors, traders, and policymakers. Our choice to study market volatility arises from its significant implications for investment strategies and economic policy, motivating us to further explore its predictive patterns and impacts.

The idea that market volatility is entirely unpredictable is challenged by empirical evidence, which identifies several consistent patterns. Volatility clustering is one such pattern, indicating that periods of high or low volatility tend to follow each other. This concept significantly influenced the creation of the ARCH and GARCH models by \citet{engle1982} and \citet{bollerslev1986}, respectively. Another key observation is the leverage effect, which shows an inverse relationship between stock prices and volatility, suggesting a complex interaction between the two \citep{black1976}. Moreover, volatility exhibits long memory, meaning market shocks can affect future volatility for an extended period, and it displays seasonal trends influenced by trading activities and economic news. These characteristics, while indicating some level of predictability, underline the complexity of financial markets \citep{andersen1998}. 

Traditional approaches to estimating market volatility have primarily centered around historical and implied volatility measures. Historical volatility is calculated based on past price fluctuations of an asset, serving as a backward-looking indicator that reflects the degree of variation in trading prices over a certain period \citep{engle1982}. Despite its utility, historical volatility is limited by its reliance on past data, failing to account for future market expectations. Implied volatility, on the other hand, is derived from the pricing of options and represents the market's forecast of a security's volatility over the life of the option \citep{black1976}. It is considered a forward-looking measure, capturing investor sentiment and expectations about future volatility. The Chicago Board Options Exchange's Volatility Index (VIX), a popular implied volatility indicator, is often referred to as the 'fear index' due to its ability to reflect market uncertainty and investor sentiment \citep{whaley2000}. While these traditional volatility estimates provide valuable insights, they are not without limitations. Historical volatility does not necessarily predict future volatility and implied volatility can be influenced by factors beyond mere expectations of future price movements, such as supply and demand dynamics for options. This has led researchers to explore more sophisticated models, such as the Generalized Autoregressive Conditional Heteroskedasticity (GARCH) model, which allows for varying volatility over time. The evolution of GARCH models has included model variations to better capture market characteristics like the introduction of the Fractionally Integrated GARCH (FIGARCH) model, which addresses the long memory characteristic of volatility, providing a more accurate representation of how past shocks can influence future volatility over extended periods \citep{baillie1996}. Additionally, the GJR-GARCH model by Glosten, Jagannathan, and Runkle integrates an asymmetry component, better capturing the leverage effect where negative market returns lead to higher subsequent volatility compared to positive returns \citep{glosten1993}.
Then, \citet{awartani2005predicting} performed a pairwise comparison of various GARCH-type models versus the GARCH(1,1) model. The findings proved the superior predictive performance of asymmetric GARCH models over the GARCH(1,1) model for both one-day ahead and longer forecast horizons. Conversely, among models not accounting for asymmetries, GARCH(1,1) showcased enhanced predictive capabilities, underlining the complex dynamics of volatility forecasting and the pivotal role of model features like asymmetry in achieving forecasting accuracy.

Recent advancements in machine learning have significantly improved the prediction accuracy of financial market volatility. Particularly, Long Short-Term Memory (LSTM) networks, a type of recurrent neural network, have shown promising results in capturing the temporal dependencies and nonlinear patterns in market data. The LSTM's ability to process data sequences makes it particularly suitable for time-series forecasting tasks, such as predicting the volatility of the S\&P 500 index. \citet{christensen2021machine} demonstrate the efficacy of LSTM models in forecasting one-day-ahead volatility of the Dow Jones Industrial Average (DJIA), comparing their performance against traditional models like the Heterogeneous AutoRegressive (HAR) model and other machine learning approaches, including regularization and tree-based algorithms. The study highlights LSTMs' superior ability to model complex dynamics in financial markets, owing to their architecture that can remember information over long periods and dynamically adjust to new information. This is particularly relevant for financial markets where volatility clustering and leverage effects present challenging forecasting conditions. 
Furthermore, a comprehensive examination of machine learning models in algorithmic trading strategies has shown that diverse approaches, such as Neural Networks, K Nearest Neighbor, Regression Trees, Random Forests, Naïve Bayes classifiers, Bayesian Generalized Linear Models, and Support Vector Machines, have been successful in forecasting market volatility across various equity indices, including the S\&P 500 \citep{grudniewicz2023}. These models were found particularly adept at utilizing technical analysis indicators to predict volatility, contributing to enhanced trading signals. The empirical evidence suggests that Linear Support Vector Machines and Bayesian Generalized Linear Models, in particular, provided the best risk-adjusted returns, thereby affirming the potential of integrating sophisticated machine learning models for financial market forecasting.
Among the neural network architectures studied, recurrent neural networks, including Long Short-Term Memory (LSTM) networks and Nonlinear Autoregressive Networks with Exogenous Inputs (NARX), stand out. Research by Bucci (2020) demonstrates that these advanced neural network models surpass traditional econometric techniques in forecasting accuracy. The enhanced capability of LSTM and NARX networks to recognize and model long-range dependencies offers a significant advantage, particularly in environments marked by high volatility, thereby positioning them as highly effective tools for predictive tasks in finance \citep{Bucci2020RealizedVF}. 
Exploring the effectiveness of machine learning in financial forecasting, Rahimikia and Poon (2020) dive into the predictive capabilities of these technologies specifically for realized volatility. Their comprehensive study assesses the performance of machine learning models using diverse data sets, including those typically associated with Heterogeneous Autoregressive (HAR) models, data from the limit order book (LOB), and news sentiment analysis. Remarkably, after training and evaluating 3.7 million models, the findings suggest that machine learning approaches, especially those managing high-dimensional data across multiple time lags, consistently surpass traditional HAR models in non-extreme volatility conditions. This indicates a superior ability of machine learning models to adapt and respond to evolving market dynamics over time, highlighting their potential to enhance forecasting accuracy in financial markets \citep{Rahimikia2020MachineLF}. 
\citet{Jia2021ForecastingVO} particularly focus on improving forecasting precision by integrating a likelihood-based loss function within deep neural network (DNN) and Long Short-Term Memory (LSTM) models. Their research contrasts traditional econometric approaches with modern machine learning techniques, demonstrating that the deep learning models equipped with this novel loss function outperform standard econometric models and traditional deep learning models using a distance loss function. This finding is critical as it confirms the potential of tailored loss functions in enhancing model performance in highly volatile market conditions, making a significant contribution to the literature on financial volatility forecasting.

Moreover, recent studies have explored the integration of GARCH models with machine learning techniques to enhance predictive accuracy. Research by \citet{kumar2020, hansen2021} on combining GARCH models with Artificial Neural Networks (ANN) and Long Short-Term Memory (LSTM) networks has shown promising results in capturing complex patterns in volatility data that traditional models may miss. These hybrid approaches leverage the time-varying volatility modeling capabilities of GARCH models and the pattern recognition strengths of machine learning algorithms, offering a powerful toolkit for forecasting market volatility. Furthermore, in the study by \citet{AmirshahiLahmiri2023}, the efficacy of integrating GARCH-type models with deep learning approaches for volatility forecasting across diverse markets was rigorously examined. This research shows how features extracted from GARCH-type models, when utilized as inputs for deep learning models like DFFNN and LSTM, significantly boost predictive performance. The integration of Long Short-Term Memory (LSTM) networks with GARCH-type models represents a significant advancement in the prediction of financial market volatility. \citet{kim2018forecasting} developed a hybrid model that combines the strengths of LSTM networks in capturing long-term dependencies and GARCH-type models in modeling time-varying volatility. Their study utilized the KOSPI 200 index data to demonstrate that the hybrid model, particularly the GEW-LSTM model combining LSTM with three GARCH-type models (GARCH, EGARCH, and EWMA), significantly outperforms traditional GARCH models and other machine learning approaches in predicting stock price index volatility. The findings underscore the potential of leveraging machine learning techniques alongside econometric models to enhance the predictive accuracy of market volatility.

The Chicago Board Options Exchange Volatility Index (VIX), often referred to as the 'fear gauge' plays a pivotal role in market volatility prediction. Fleming, Ostdiek, and Whaley (1995) laid the groundwork by illustrating the dynamic relationship between the VIX and market returns, highlighting its utility in forecasting market volatility \citep{fleming1995}. Building on this foundation, \citet{whaley2000} underscored the significance of the VIX as a leading indicator for market stress and investor uncertainty, demonstrating its predictive capacity for future market movements. Further exploring the integration of VIX data into volatility models, \citet{todorov2010} enhanced the prediction of volatility by incorporating market expectations, thereby providing a more nuanced understanding of market dynamics. These studies collectively affirm the VIX's crucial role in providing insights into future market volatility, serving as an invaluable tool for investors, portfolio managers, and policymakers in navigating financial markets. Based on the recent study, \citet{Bhandari2022PredictingSM} leveraged the power of artificial intelligence and machine learning, specifically utilizing a Long Short-Term Memory (LSTM) network, to forecast the S\&P 500 index's next-day closing prices. They meticulously selected a combination of nine predictors, such as market data, macroeconomic data, and technical indicators to construct a holistic view of the stock market's dynamics. Their research compared single-layer and multilayer LSTM models across various performance metrics, concluding that single-layer models offer superior accuracy and fit. Notably, the VIX was included among the predictors, underscoring its significance in capturing market volatility insights. 

Considering the pivotal role of volatility in financial decision-making, this research aims to study the efficacy of standard econometric models and contemporary machine learning techniques in predicting the volatility of the S\&P 500 index. The integration of traditional models with advanced predictive techniques allows for a comprehensive examination of volatility's dynamic nature. Additionally, this study explores the application of the Local Interpretable Model-agnostic Explanations (LIME) technique on the best-performing model, aiming to enhance the interpretability of machine learning predictions in financial markets, thereby bridging the gap between complex model outputs and actionable financial insights.
Hence, guided by the insights from previous studies, our research contrasts the traditional GARCH model with complex LSTM models, which utilize a diverse array of predictors as outlined in Table\ref{tab:variables}, to forecast the S\&P 500's volatility. \\

\section{Modelling Setup}
\label{sec:data}
This research incorporates a multifaceted approach to estimate the volatility of S\&P 500, employing time series as well as a neural network model with a variety of data inputs to capture the complexity and nuances of market behavior. This includes the analysis of S\&P 500 log returns, which serve as a fundamental gauge of market movement; the examination of S\&P 500 lagged volatility, providing insights into historical volatility patterns; the utilization of GARCH model predictions, offering forward-looking estimates of S\&P 500 volatility for a given day; and the incorporation of VIX prices, known as the market's 'fear gauge' to assess expected market volatility.

\subsection{Data inputs}
\label{sec:data_description}

The analysis within this study is predicated upon the daily data of the S\&P 500, encompassing 6032 data points spanning from January 3, 2000, to December 21, 2023, sourced from Yahoo Finance. This dataset serves as the foundation for extracting pivotal variables critical for the analysis and predictive modeling of financial market behaviors, as highlighted in Table~\ref{tab:variables}. Specifically, the study employs two principal variables derived from the S\&P 500 data: Log Returns and Lagged Volatility.

\begin{figure}[H]
    \centering
    \begin{minipage}[t]{0.48\textwidth} 
        \caption{S\&P 500 Log Returns Distribution}
        \includegraphics[width=\linewidth]{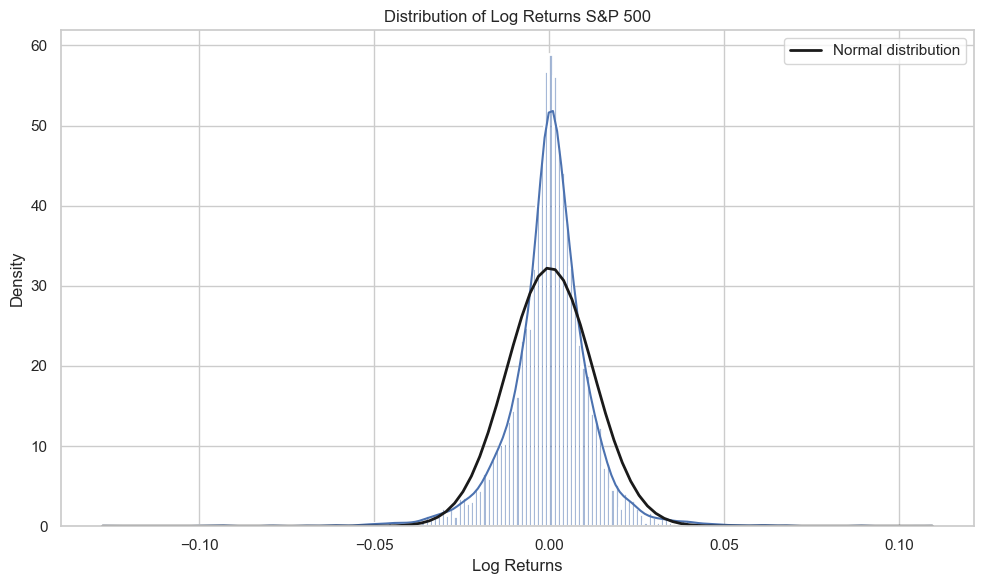}
        \label{fig:log_returns_dist}
        \caption*{\footnotesize Note: This figure compares the distribution of the S\&P 500 log returns (period from January 3, 2000, to December 21, 2023) with normal distribution.}
    \end{minipage}\hfill
    \begin{minipage}[t]{0.5\textwidth} 
        \caption{S\&P 500 Log Returns Over Time}
        \includegraphics[width=\linewidth]{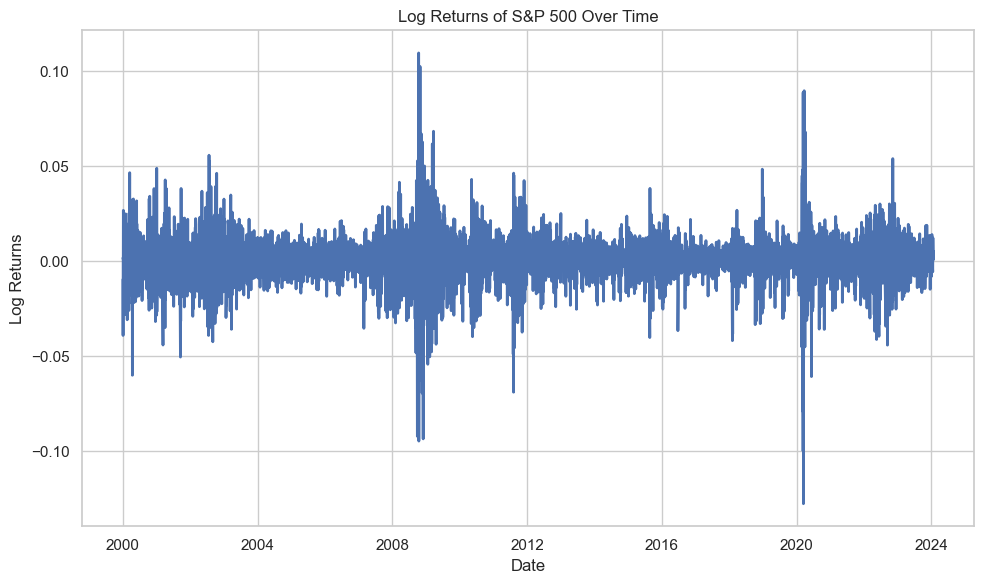}
        \label{fig:log_returns_sp500}
        \caption*{\footnotesize Note: This figure presents the time series of log-returns of the S\&P 500, mapping the return fluctuations over time. It serves to indicate the volatility and the temporal patterns in the log returns across the observed period from January 3, 2000, to December 21, 2023.}
    \end{minipage}
\end{figure}

\begin{figure}[H]
    \centering
    \begin{minipage}[t]{0.48\textwidth} 
        \caption{S\&P 500 Volatility Distribution}
        \includegraphics[width=\linewidth]{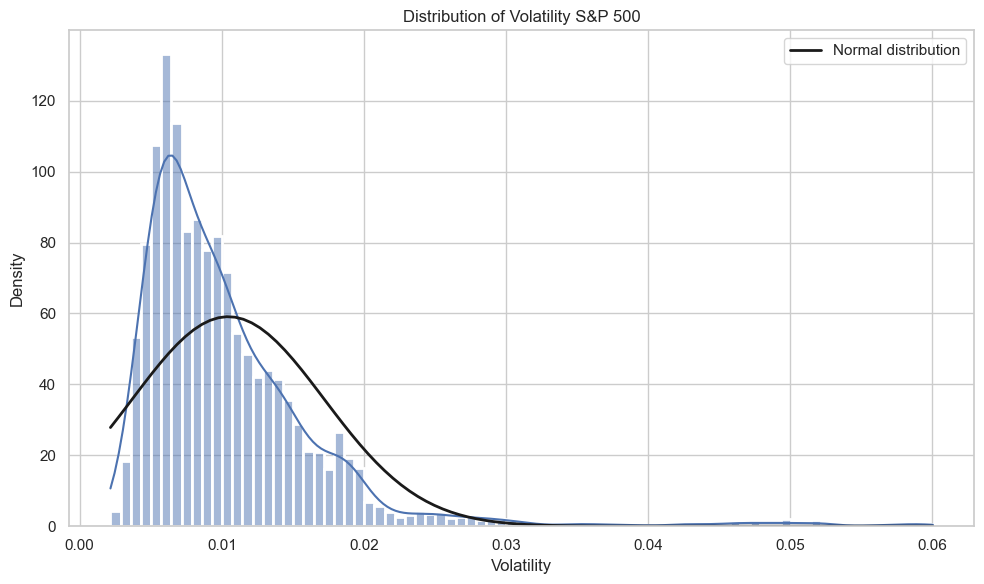}
        \label{fig:volatility_dist}
        \caption*{\footnotesize Note: This figure compares the distribution of the S\&P 500 volatility (period from January 3, 2000, to December 21, 2023) with normal distribution.}
    \end{minipage}\hfill
    \begin{minipage}[t]{0.5\textwidth} 
        \caption{S\&P 500 Volatility Over Time}
        \includegraphics[width=\linewidth]{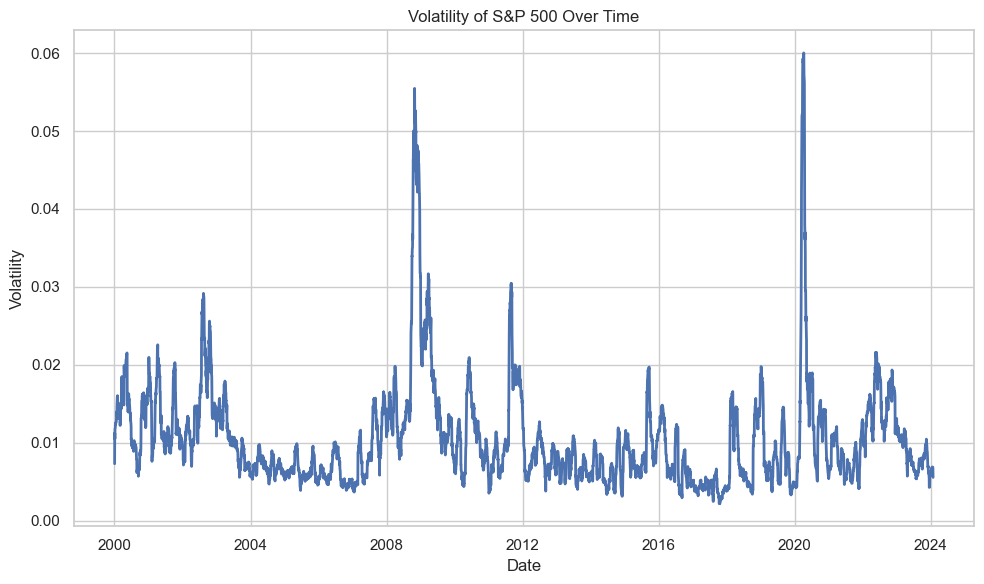}
        \label{fig:volatility}
        \caption*{\footnotesize Note: This figure presents the time series of the volatility of the S\&P 500, mapping the fluctuations over time. It serves to indicate the changes in volatility across the observed period from January 3, 2000, to December 21, 2023.}
    \end{minipage}
\end{figure}

Log returns, illustrated in Figure~\ref{fig:log_returns_dist} and Figure~\ref{fig:log_returns_sp500}, formulated in Equation~\ref{eq:log_returns}, compute the logarithmic difference between consecutive closing prices of the S\&P 500. This metric is preferred in financial analyses for its capability to normalize and facilitate comparison of price movements over time. Conversely, (Lagged) Volatility, shown in Figure~\ref{fig:volatility_dist} and Figure~\ref{fig:volatility}, outlined in Equation~\ref{eq:lagged_volatility}, captures historical volatility trends within the market, offering insights into past market dynamics and serving as a predictor of potential future volatility and market sentiment.

Further enhancing the study's analytical depth, GARCH volatility predictions are integrated as inputs into the LSTM-GARCH model, as discussed in Section~\ref{sec:modelling_approaches_garch}. The inclusion of GARCH-derived volatility predictions allows the model to emulate the temporal evolution of market uncertainty, thereby crafting a prediction framework that accounts for financial markets' intrinsic volatility.

\begin{figure}[H]
    \centering
    \begin{minipage}[t]{0.48\textwidth} 
        \caption{VIX Distribution}
        \includegraphics[width=\linewidth]{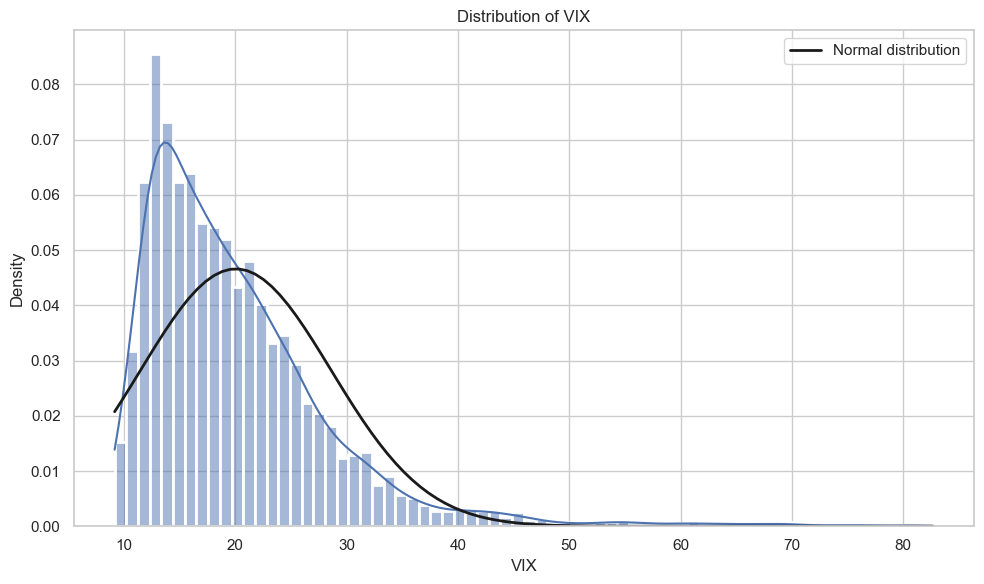}
        \label{fig:vix_dist}
        \caption*{\footnotesize Note: This figure compares the distribution of VIX (period from January 3, 2000 to December 21, 2023) with normal distribution.}
    \end{minipage}\hfill
    \begin{minipage}[t]{0.5\textwidth} 
        \caption{VIX Over Time}
        \includegraphics[width=\linewidth]{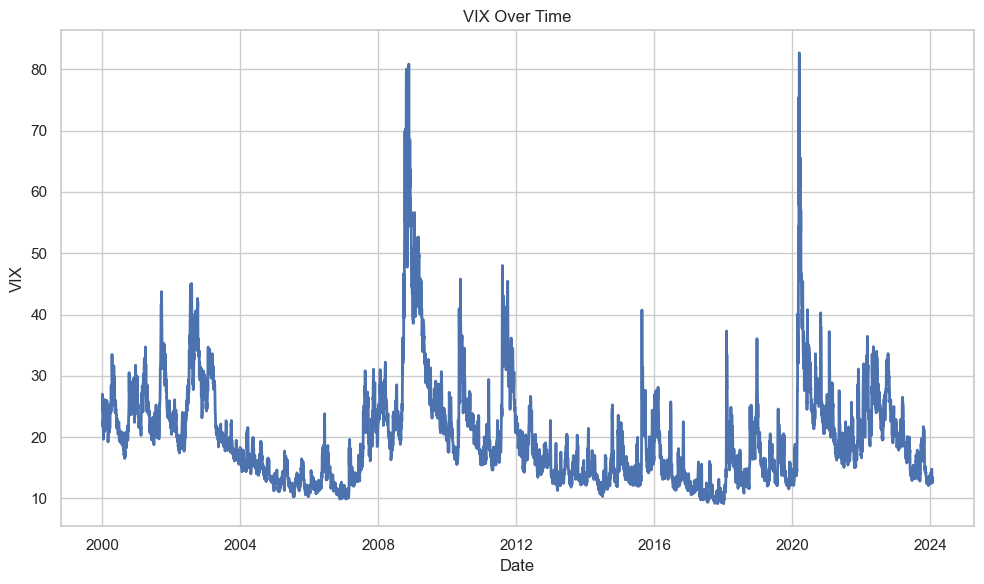}
        \label{fig:vix}
        \caption*{\footnotesize Note: This figure presents the time series of VIX across the observed period from January 3, 2000 to December 21, 2023.}
    \end{minipage}
\end{figure}

Additionally, the study leverages the VIX index (Figure~\ref{fig:vix_dist} and Figure~\ref{fig:vix}) to gain insights into investor sentiment and anticipated market volatility. As elaborated in Section~\ref{sec:modelling_approaches_lstm_garch_vix}, the VIX input infuses the models with a prospective view on volatility, derived from S\&P 500 index options, hence introducing a forward-looking element to the predictive analyses. This integration ensures the models not only rely on historical data but also consider the market's current expectations, enhancing the precision and relevance of the forecasts.

\begin{table}[H]
\centering
\caption{Summary of Variables Used in the Models}
\label{tab:variables}
\begin{tabular}{p{2cm} p{7cm} p{5.5cm}}
\hline
\textbf{Group} & \textbf{Variable} & \textbf{Used in Model} \\
\hline
S\&P 500 & Log Returns  & \makecell[tl]{LSTM, \\ LSTM-GARCH, \\ LSTM-GARCH with VIX input} \\
\hline
S\&P 500 & Lagged Volatility & \makecell[tl]{LSTM, \\ LSTM-GARCH, \\ LSTM-GARCH with VIX input} \\
\hline
GARCH & GARCH volatility estimates & \makecell[tl]{LSTM-GARCH \\ LSTM-GARCH with VIX input} \\
\hline
VIX & VIX implied volatility  & \makecell[tl]{LSTM-GARCH with VIX input} \\
\hline
\end{tabular}
\caption*{\footnotesize Note: This table enumerates all the variables utilized across the four models analyzed in this study: log returns, lagged volatility, GARCH volatility estimates, and VIX implied volatility as discussed above in Section \ref{sec:data_description}. The log returns and lagged volatility are derived from the S\&P 500 dataset. GARCH volatility estimates are calculated as described in Section \ref{sec:modelling_approaches_garch}. VIX index close prices are used as the VIX implied volatility variable. }
\end{table}

Table \ref{tab:descriptive_stats} presents the descriptive statistics for close prices and log returns. The normality of a distribution is often assessed by the proximity of its mean to the median. The discrepancy between the mean and the median indicates a potential skew in the distribution of close prices. On the other hand, a closer mean to the median for log returns suggests a distribution with a tendency toward normality. Extreme values, as indicated by the distance between the maximum and minimum as well as the 25th and 75th percentiles, imply a distribution with heavy tails, thus higher kurtosis. The mean log return is \(0.0002\), indicating a marginal average daily increase in asset price, with a standard deviation of \(0.0124\), which reflects the extent of return volatility. The range of log returns spans from a minimum of \(-0.1277\) to a maximum of \(0.1096\), showcasing significant fluctuations in asset price movements. Quartile values further delineate the distribution, with 25\%, 50\% (median), and 75\% percentiles located at \(-0.0049\), \(0.0006\), and \(0.0059\), respectively.

The log returns distribution's normality was challenged through the Shapiro-Wilk test, which yielded a statistic of \(0.9005\) and a p-value of \(0.0\), indicating a slight deviation from a normal distribution. These results are confirmed by measures of skewness and kurtosis, with values of \(-0.37859\) and \(10.29295\) respectively, indicating a left-skewed distribution and a leptokurtic shape, which are characteristic of financial return data with its pronounced tails and peak.

\begin{table}[H]
\centering
\caption{Descriptive Statistics of Close Prices and Log Returns of S\&P500}
\label{tab:descriptive_stats}
\begin{tabular}{p{6cm} p{3cm} p{3cm}}
\hline
\textbf{Statistic} & \textbf{Close Prices} & \textbf{Log Returns} \\
\hline
Count & 6014 & 6054 \\
Mean & 1985.29 & 0.0002 \\
Standard Deviation & 1072.75 & 0.0124 \\
Minimum & 676.53 & -0.1277 \\
25\% Percentile & 1191.15 & -0.0049 \\
50\% Percentile (Median) & 1462.46 & 0.0006 \\
75\% Percentile & 2624.44 & 0.0059 \\
Maximum & 4894.16 & 0.1096 \\
\hline
Shapiro-Wilk Statistic &  0.66 & 0.90 \\
Shapiro-Wilk p-value &  0.00 & 0.00 \\  
\hline
\end{tabular}

\caption*{\footnotesize Note: Panel A presents the descriptive statistics for close prices, showing a discrepancy between the mean and the median, indicating a potential skew in the distribution. Panel B presents the descriptive statistics for log returns. The closer mean to median in log returns (compared to close prices) suggests a distribution with tendencies toward normality, despite significant fluctuations and heavy tails. The Shapiro-Wilk test results confirm a deviation from normal distribution, indicated by the skewness and kurtosis measures. However, such characteristics are typical for financial returns data.}
\end{table}

\vspace{-10pt}

\vspace{-10pt}

\vspace{-10pt}

\subsection{Estimation of log returns and volatility}
\label{sec:estimation_of_log_returns_and_volatility}
The log returns of S\&P 500 on a day $t$ are calculated as a natural logarithm of the ratio of the closing price on day $t$ to the closing price on the previous day $(t-1)$.

\begin{equation}
\text{Log\_Returns}_t = \log\left(\frac{\text{Close}_t}{\text{Close}_{t-1}}\right)
\label{eq:log_returns}
\end{equation}

Then, the historical volatility estimate on day $t$ is calculated as the standard deviation of log returns over a specified rolling window. 

\begin{equation}
\text{Volatility}_t = \sqrt{\frac{1}{N-1} \sum_{i=t-N+1}^{t} (\text{Log\_Returns}_t - \mu)^2}
\label{eq:volatilty}
\end{equation}

where:
\begin{itemize}
    \item $N$ denotes the 22 trading days window, a model time frame commonly approximating a trading month,
    \item $\mu$ is the mean average of the log returns over the rolling window. 
\end{itemize}

Next, the lagged volatility for a given day is calculated by shifting the volatility values by a specified number of lag days. For a lag of 1 day, the calculation can be represented as follows:

\begin{equation}
\text{Lagged\_Volatility}_{t} = \text{Volatility}_{t-1}
\label{eq:lagged_volatility}
\end{equation}

where:
\begin{itemize}
    \item $\text{Volatility}_{t-1}$ represents the volatility on the previous day $(t-1)$.
    \item $\text{Lagged\_Volatility}_{t}$ represents the lagged volatility value assigned to day $t$.
\end{itemize}

\subsection{Data standardization}
\label{sec:data_standarization}
Since standardization is a key step before inputting data into the LSTM models, the Min-Max scaling method was used to normalize the raw input data. This normalization process is applied dynamically to each prediction window within our time series dataset, ensuring that our model is trained on data that reflects the most recent conditions. The formula for Min-Max scaling is given by: 
\begin{equation}
X_{\text{scaled}} = \frac{X - X_{\min}}{X_{\max} - X_{\min}}
\label{eq:minmax_scaling}
\end{equation}

where: $X$ is the original value, $X_{\min}$ is the minimum value in the feature column, $X_{\max}$ is the maximum value in the feature column and $X_{\text{scaled}}$ is the resulting normalized value.

The scalers are fitted on the current training data segment. This process recalculates the minimum and maximum values used for scaling, based on the training data available up to the current prediction point. After fitting, the training, validation, and subsequent test datasets are transformed using these dynamically adjusted scalers. The model is either trained or loaded from previous states, and predictions are made on the normalized test data. Post-prediction, the output is inversely transformed to revert it back to the original data scale for accurate error calculation and comparison.

\subsection{Error Metrics}

Error metrics are crucial for evaluating the performance of predictive models, providing quantitative measures of the accuracy of the predictions. Two commonly used error metrics are the Mean Absolute Error (MAE) and the Root Mean Squared Error (RMSE). 

\subsubsection{Mean Absolute Error (MAE)}

The Mean Absolute Error (MAE) is a measure of the average magnitude of the errors in a set of predictions, without considering their direction. It calculates the average of the absolute differences between the predicted values and the actual values. The MAE is given by the equation:

\begin{equation}
MAE = \frac{1}{n} \sum_{i=1}^{n} |y_i - \hat{y}_i|
\label{eq:mae}
\end{equation}

where:
\begin{itemize}
    \item $n$ is the number of observations,
    \item $y_i$ is the actual value for the $i$th observation,
    \item $\hat{y}_i$ is the predicted value for the $i$th observation,
    \item $|\cdot|$ denotes the absolute value.
\end{itemize}

The MAE provides a straightforward measure of prediction error magnitude, with lower values indicating better model performance.

\subsubsection{Root Mean Squared Error (RMSE)}

The Root Mean Squared Error (RMSE) quantifies the square root of the average of the squared differences between the predicted and actual values. Unlike MAE, RMSE gives more weight to larger errors due to the squaring of the error terms. The RMSE is defined as:

\begin{equation}
RMSE = \sqrt{\frac{1}{n} \sum_{i=1}^{n} (y_i - \hat{y}_i)^2}
\label{eq:rmse}
\end{equation}

where the variables are as previously defined. RMSE is sensitive to outliers and tends to penalize large errors more heavily than MAE. Lower RMSE values denote higher accuracy of the predictive model. \\

Both MAE and RMSE are scale-dependent metrics, meaning their values are influenced by the scale of the data. They are best used for comparative purposes—to compare the performance of different models or model configurations on the same dataset. \\

\section{Modelling approaches}
\label{sec:modelling_approaches}

\subsection{GARCH}
\label{sec:modelling_approaches_garch}

The Generalized Autoregressive Conditional Heteroskedasticity (GARCH) model, introduced by Bollerslev (1986), extends the foundational Autoregressive Conditional Heteroskedasticity (ARCH) model proposed by Engle (1982). The GARCH model articulates the conditional variance as a linear function of past squared values of the series, enabling it to adeptly capture the volatility clustering phenomenon prevalent in financial markets. Such a pattern, characterized by periods of high volatility followed by similar periods, highlights the GARCH model's efficacy in financial time series analysis \citep{bollerslev1986, engle1982}.

\subsubsection{Model Selection and Parameterization}
\label{sec:model_selection_and_parameterization_garch}

As described by \citep{francq2010garch}, the GARCH model is defined by its two parameters, \( p \) and \( q \), which govern the number of lagged conditional variances and squared shocks, respectively, included in the volatility equation. Specifically, \( q \) represents the order of the ARCH terms, capturing the short-term volatility due to recent shocks, while \( p \) represents the order of the GARCH terms, encapsulating the volatility's persistence across time. The model's conditional variance \( \sigma_t^2 \) is expressed through these parameters as follows:

\begin{equation}
\sigma_t^2 = \omega + \sum_{i=1}^{q} \alpha_i \epsilon_{t-i}^2 + \sum_{j=1}^{p} \beta_j \sigma_{t-j}^2
\end{equation}

where \( \omega \) is the intercept term, \( \alpha_i \) are the coefficients of the ARCH component, and \( \beta_j \) are the coefficients of the GARCH component. The backshift operator \( B \) simplifies the representation, allowing for the compact expression of the model as:

\begin{equation}
\sigma_t^2 = \omega + \alpha(B)\epsilon_t^2 + \beta(B)\sigma_t^2
\end{equation}

with the polynomials \( \alpha(B) \) and \( \beta(B) \) defined as \( \alpha(B) = \sum_{i=1}^{q} \alpha_i B^i \) and \( \beta(B) = \sum_{j=1}^{p} \beta_j B^j \), respectively. In the special case where \( \beta(z) = 0 \), the GARCH model simplifies to an ARCH model with no autoregressive conditional variances. \\

To identify the optimal parameters for the \(GARCH(p, q)\) model in forecasting financial market volatility, an exhaustive exploration of parameter combinations was undertaken. Parameters \(p\) and \(q\) were varied within the range from 0 to 4 in a systematic loop. This analysis pinpointed the GARCH(2,2) model as the optimal choice, marked by the lowest Akaike Information Criterion (AIC) score of 16750.227, hence balancing fitting accuracy with model simplicity effectively. The Akaike Information Criterion is a measure used for model selection where lower values suggest a better model fit, adjusted for the number of parameters used, as introduced by \cite{akaike1974new}. It is calculated as \( \text{AIC} = 2k - 2\ln(L) \), where \( k \) is the number of parameters and \( L \) is the maximum likelihood of the model. The model is mathematically expressed as:

\begin{equation}
\sigma_t^2 = \omega + \alpha_1 \epsilon_{t-1}^2 + \alpha_2 \epsilon_{t-2}^2 + \beta_1 \sigma_{t-1}^2 + \beta_2 \sigma_{t-2}^2
\label{eq:garch22}
\end{equation}

\subsubsection{GARCH Volatility Forecasting Methodology}
\label{sec:volatility_forecasting-methodology_garch}

The forecast method uses a GARCH(2,2) model to predict future changes in the S\&P 500 index's volatility. 

The analysis initiates with S\&P 500 index data, particularly focusing on daily log returns (as per Equation ~\ref{eq:log_returns}) for their advantageous properties in financial modeling \cite{tsay2010analysis}. Log returns are preferred for their additivity over time, which allows for straightforward aggregation of returns across time intervals. Additionally, log returns are often more homoskedastic, providing a more consistent variance that is conducive to the assumptions underlying various volatility models. The symmetry and normalization features of log returns, along with their tendency to align closer to a normal distribution, further validate their use in predictive financial models.

Following the model's parameterization, the dataset was segmented into an initial training period (January 1, 1985, to January 2, 2000) and a testing period (January 3, 2000, to December 21, 2023).  The prediction approach involves updating the training set with the most recent data at each step, thereby allowing for the prediction of the subsequent day's (\(t+1\)) volatility. At every iteration, the model is trained on the current dataset to forecast the volatility for the next day. This process not only ensures that the model remains dynamically aligned with the latest market trends but also enables a robust evaluation of the model's predictive accuracy in a real-time context.

For the GARCH model to deliver trustworthy estimates and predictions, the time series data need to be stationary, which means its statistical properties, like mean and variance, don't change over time. We applied the Augmented Dickey-Fuller (ADF) test to the scaled log returns to check for stationarity. The ADF test's null hypothesis suggests that the time series has a unit root and is non-stationary. With a test statistic of \(-33.4127\) and a p-value of \(0.0\), we have strong evidence to reject the null hypothesis, confirming that the data is indeed stationary.

\subsubsection{GARCH Results}
\label{sec:results_garch}

\begin{figure}[H]
    \centering
    \caption{GARCH Out of Sample prediction}
    \includegraphics[width=1\linewidth]{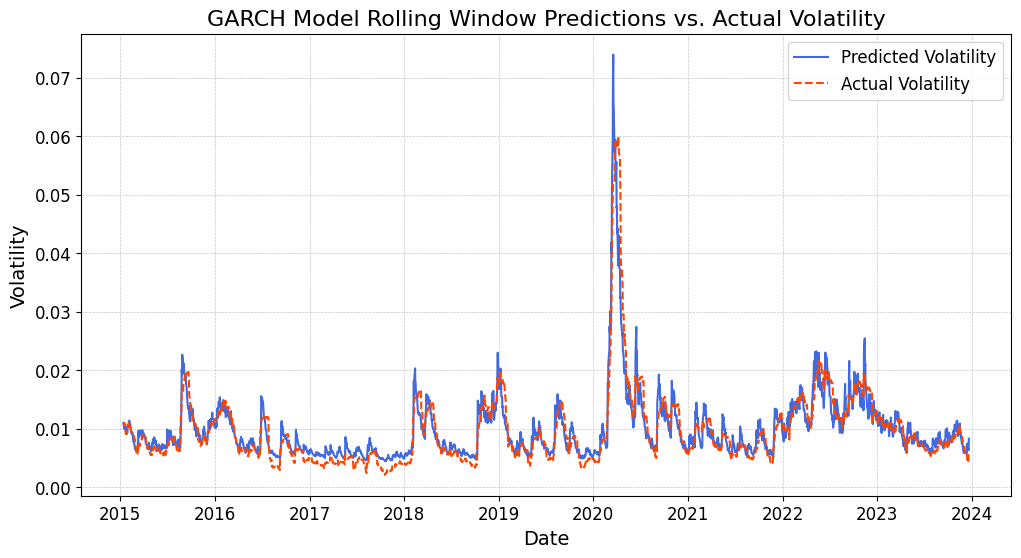}
    \label{fig:garch_results}
    \caption*{\footnotesize Note: This figure presents the GARCH Out of Sample prediction over the period February 13, 2015 to December 21, 2023.}
\end{figure}

The performance of the GARCH model's volatility predictions can be visually assessed in Figure~\ref{fig:garch_results}, where the model's predictions are plotted alongside the actual market volatility over a testing period from February 13, 2015, to December 21, 2023. Generally, the predicted volatility, represented by the solid blue line, tends to be higher than the actual market volatility, shown with a dashed line, across most periods. This trend of overestimation is particularly pronounced during periods of market stress, such as the significant spike observed around 2020. While this conservative bias in volatility forecasting could be beneficial for risk management by preparing for potential higher volatility, it also highlights an overarching tendency of the model to overestimate, especially evident in capturing extreme market movements, suggesting areas for improvement in its predictive accuracy.

A quantitative assessment of the model's out-of-sample prediction accuracy is presented in Table~\ref{tab:garch_metrics}. The MAE was \num{1.56e-3} and the RMSE was \num{2.39e-3}. These error metrics are relatively low, which corroborates the visual assessment that the GARCH model predictions are in close agreement with the observed volatility, although the model's performance during extreme market conditions could be further examined and improved.

\begin{table}[H]
\centering
\caption{GARCH Out-of-Sample Error Metrics}
\label{tab:garch_metrics}
\begin{tabular}{
  >{\raggedright\arraybackslash}p{8cm}
  S[table-format=1.3e-1]
  S[table-format=1.3e-1]
  }
\hline
\textbf{Model} & {\textbf{MAE}} & {\textbf{RMSE}} \\
\hline
GARCH & 1.56e-3 & 2.39e-3 \\
\hline
\end{tabular}
\caption*{\footnotesize Note: This table presents the GARCH Out of Sample Error Metrics, for walk-forward predictions on a t+1 basis over the period from February 13, 2015 to December 21, 2023.}
\end{table}

\subsection{LSTM}
\label{sec:modelling_approaches_lstm}

Long Short-Term Memory (LSTM) networks, introduced by \cite{hochreiter1997long}, mark a pivotal advancement in recurrent neural network (RNN) architectures, enabling the learning of order dependencies in sequence prediction problems. Unlike conventional time-series forecasting methods, LSTMs excel in capturing and retaining information across extended sequences of data. This capability makes them particularly well-suited for analyzing financial time series data, which is often characterized by noise and non-stationarity.

LSTMs stand out for their memory cells, which capture long-term dependencies and temporal dynamics inherent in financial market volatility. Unlike traditional RNNs, LSTMs have a complex architecture that includes multiple gates to control the flow of information, addressing some of the key challenges that RNNs face, such as the difficulty in making use of distant information and the vanishing gradient problem.

In the LSTM cell architecture the input at the current timestep, denoted as \(x_t\), and the hidden state from the preceding timestep, \(h_{t-1}\), are aggregated and fed into three critical gates: the forget gate, the input gate, and the output gate. These gates regulate the flow of information through the cell by selectively updating the cell state and the hidden state based on the inputs they receive.

\begin{itemize}
    \item \textbf{Forget Gate:} Here, the activation function is given by \(f_t = \sigma(U_f \cdot x_t + V_f \cdot h_{t-1} + b_f)\), where \(f_t\) signifies the activation output of the forget gate, and \(b_f\), \(U_f\), and \(V_f\) represent the biases, input weights, and recurrent weights, respectively. This sigmoid function, outputting values between 0 and 1, determines the degree to which information from the previous cell state, \(C_{t-1}\), is retained or discarded in the new cell state, with the updated cell state being \(C'_t = f_t \cdot C_{t-1}\).
    
    \item \textbf{Input Gate:} This gate determines the new information to be incorporated into the cell state from the current input and the previous hidden state. The process involves two steps: calculating \(i_t = \sigma(U_i \cdot x_t + V_i \cdot h_{t-1} + b_i)\) to decide which parts of the new information are significant, and then forming a new candidate cell state, \(C_t^+ = \tanh(U_c \cdot x_t + V_c \cdot h_{t-1} + b_c)\). The cell state is updated as \(C_t = C'_t + i_t \cdot C_t^+\), blending the old and new information.
    
    \item \textbf{Output Gate:} Employing a sigmoid function, \(o_t = \sigma(U_o \cdot x_t + V_o \cdot h_{t-1} + b_o)\), this gate influences the update of the hidden state for the current timestep. It uses inputs from the previous hidden state and the current input to determine how the current cell state, \(C_t\), influences the new hidden state, \(h_t = o_t \cdot \tanh(C_t)\), thus deciding what portion of the cell state is conveyed in the output.
\end{itemize}

\begin{figure}[H]
    \centering
    \caption{Long Short-Term Memory Model Architecture}
    \includegraphics[width=1\linewidth]{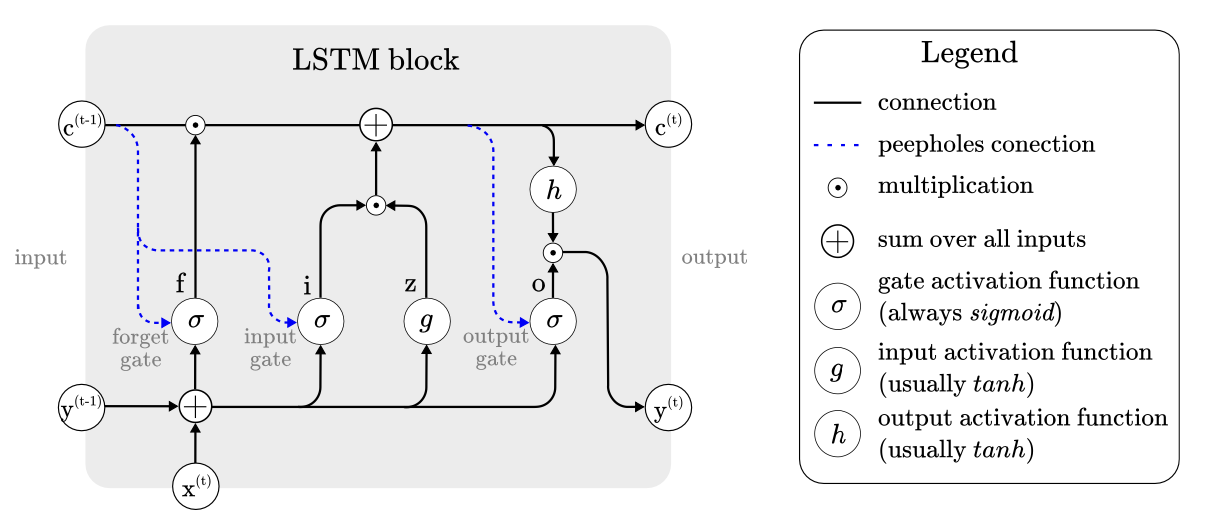}
    \caption*{\footnotesize Note: Illustration from "A Review on the Long Short-Term Memory Model" by Van Houdt et al. (2020), showcasing the architecture of an LSTM network. Reprinted under © Springer Nature B.V. 2020.}
    \label{fig:lstm_architecture}
\end{figure}

\subsubsection{Model Selection and Parameterization}
\label{sec:model_selection_and_parameterization_lstm}

To enhance the LSTM model's predictive accuracy for financial market volatility, a comprehensive hyperparameter tuning process was conducted on a dataset spanning fifteen years, from 2000 to 2015. The Keras Tuner library in Python, implementing a RandomSearch strategy, was used for this optimization. This methodical approach involved setting the search to execute 50 trials, each with 3 executions, to ensure the reliability of the results. Additionally, EarlyStopping was employed to prevent model overfitting, optimizing the model's performance while safeguarding against the loss of generalizability.

The tuning process included: the number of layers (1, 2, or 3), the number of neurons in the LSTM layers (32, 64, or 128), activation functions for the LSTM layers ('tanh' and 'relu' activations), dropout rates (varied from 0.0 to 0.3 in increments of 0.1), learning rates for the Adam optimizer (0.01, 0.001, 0.0001), loss function (MSE or MAE).

\begin{table}[H]
\centering
\caption{Hyperparameter search space of LSTM model}
\label{tab:lstm_tuning}
\begin{tabular}{p{12cm}l}
\hline
\textbf{Hyperparameter} & \textbf{Value Range} \\
\hline
Number of Layers in the LSTM network & 1, 2, 3 \\
Number of Neurons in the LSTM layers & 32, 64, 128 \\
Activation Functions for the LSTM layers & 'tanh', 'relu' \\
Dropout Rates & 0.0, 0.1, 0.2, 0.3 \\
Learning Rates for the Adam Optimizer & 0.01, 0.001, 0.0001 \\
Loss function & MSE, MAE \\
\hline
\end{tabular}
\caption*{\footnotesize Note: This table presents hyperparameters search space for LSTM model architecture. RandomSearch was used with max trials 50, execution per trial 3, epochs 50, and with an early stopping mechanism with patience 10. The search was performed on the 15 years of data, from January 2000 to January 2015.}
\end{table}

Following the tuning phase, the architecture selected for the LSTM network comprised two LSTM layers, accompanied by a dense output layer activated by a ReLU function. Given the complexity of LSTM models and the considerable duration required for their execution, the research was constrained to employing a single, optimally tuned LSTM model for the entire testing period due to time and computing power limitations. Table \ref{tab:lstm_final_config} shows the final configuration of hyperparameters for the LSTM model.

\begin{table}[H]
\centering
\caption{Final Hyperparameter Configuration of LSTM Model}
\label{tab:lstm_final_config}
\begin{tabular}{p{12cm}l}
\hline
\textbf{Hyperparameter} & \textbf{Configuration} \\
\hline
Number of Layers in the LSTM network & 2 \\
Number of neurons in the first hidden layer  & 128 \\
Number of neurons in the second hidden layer  & 128 \\
Activation function for the first hidden layer & 'tanh' \\
Activation function for the second hidden layer & 'tanh' \\
Dropout rate after first hidden layer & 0.1 \\
Dropout rate after second hidden layer & 0.1 \\
Learning rate for the Adam optimizer & 0.001 \\
Loss function & MSE \\
Epochs & 100 \\
Batch size & 64 \\
\hline
\end{tabular}
\caption*{\footnotesize Note: This table presents the final LSTM architecture that was selected during hyperparameters tuning on the 15 years of data, from January 2000 to January 2015.}
\end{table}

\subsubsection{LSTM Volatility Forecasting Methodology}
\label{sec:volatility_forecasting_methodology_lstm}

The tuned LSTM architecture includes two LSTM layers with 128 neurons each. The first layer includes a 'tanh' activation function and a recurrent dropout of 0.1, while the second layer does not return sequences and also employs a 'tanh' activation. After each LSTM layer, a dropout of 0.1 is applied to prevent overfitting. The network concludes with a dense output layer activated by ReLU, known for its ability to maintain non-linearity in output while ensuring positive predictions, an essential characteristic for forecasting metrics like volatility that are inherently positive. 

The dataset utilized for forecasting volatility comprises log returns (Equation \ref{eq:log_returns}) and lagged volatility (Equation \ref{eq:lagged_volatility}) measures. Volatility, as defined in Equation ~\ref{eq:volatilty}, is set as the target variable.
Data preprocessing involves scaling the feature variables with a Min-Max Scaler (Equation \ref{eq:minmax_scaling}). The scaled features and target are then structured into sequences to be LSTM-friendly, with a lookback period of 22-time steps. The 22-day lookback period reflects the approximate number of trading days in a month, considering financial markets usually operate five days a week. This standard is used to align model inputs with monthly market cycles effectively. 
To enhance reproducibility across computational environments, random seeds were consistently set across various libraries. The model was compiled using the Adam optimizer combined with a mean squared error loss function. This configuration is particularly well-suited for the regression nature of volatility forecasting, aiming to minimize discrepancies between actual and predicted values.

To mitigate the risk of overtraining- a common issue where the model overly specializes on the training data to the detriment of its generalization to unseen data- an early stopping mechanism was integrated into the training routine. This technique monitors the model's performance on a validation set, stopping the training process if there is no improvement in predictive accuracy for a predefined number of epochs, set in this case to 10. This parameter, known as the patience parameter, strikes a balance between sufficient model training and the prevention of overfitting, ensuring that the model remains generalizable to new market conditions.

\begin{figure}[H]
    \centering
    \caption{Walk Forward Validation}
    \includegraphics[width=1\linewidth]{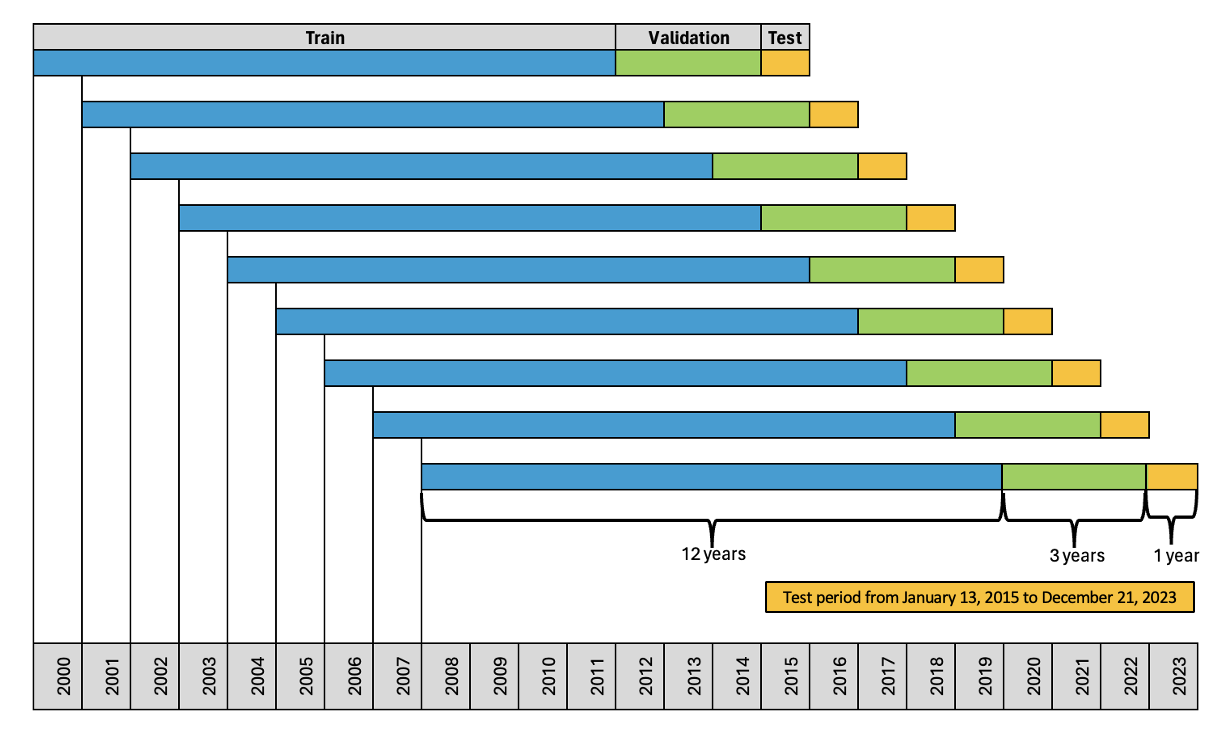}
    \label{fig:walk_forward}
    \caption*{\footnotesize Note: Initially, the training dataset consisted of 12 years, or 3,024 days (12 $\times$ 252 days), and the validation set included 756 days (3 $\times$ 252 days). The model was refitted every 252 days, which is the size of one walk-forward window. New observations were incrementally added to the datasets to ensure currency without full retraining. Predictions were generated for the next day (\(t+1\)).
}
\end{figure}

A walk-forward methodology (Figure \ref{fig:walk_forward}) was employed as the foundation of the model training process. Initially, the training dataset consisted of 12 years of data, totaling 3,024 days (12 $\times$ 252 days), and the initial validation set was comprised of 756 days (3 $\times$ 252 days). The model was refitted every 252 observations, which is a common practice reflecting the typical number of trading days in a year. This approach ensures that the model's weights are consistently updated with the most recent market data, enhancing its responsiveness to new information. Furthermore, to integrate new observations without retraining from scratch, the model incrementally adds them, maintaining an up-to-date dataset. Predictions are then generated for the immediate future time step (\(t+1\)), aligning the forecasting process with practical, short-term market predictions. This process not only maintained the relevance of the model to current market conditions but also enhanced the robustness of the forecasts by continuously integrating new information into the training and validation phases.

The model's predictive performance is assessed through two key metrics: the MAE and the RMSE. Both metrics offer a straightforward evaluation of prediction quality by averaging the discrepancies between predicted and actual values. Specifically, MAE calculates the average of the absolute differences, providing a clear measure of the magnitude of prediction errors, while RMSE squares these differences before averaging, thus giving more weight to larger errors and offering insight into the variance of the predictions.

\subsubsection{LSTM Results}
\label{sec:results_lstm}

\begin{figure}[H]
    \centering
    \caption{LSTM Out-of-sample prediction}
    \includegraphics[width=1\linewidth]{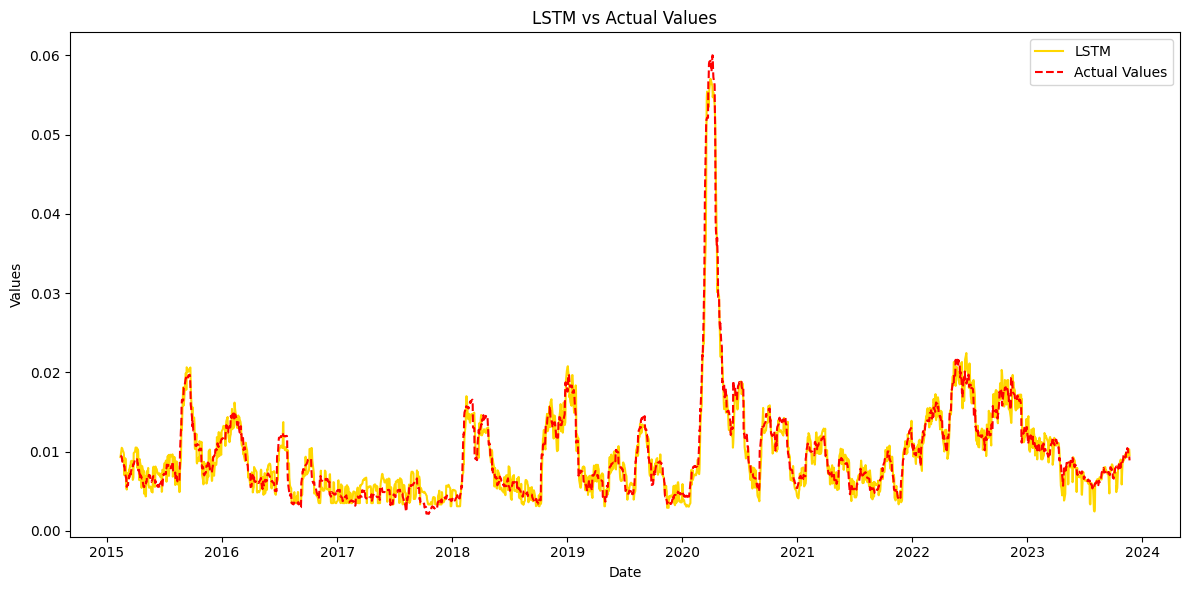}
    \label{fig:lstm8_results}
    \caption*{\footnotesize Note: This figure presents the LSTM Out of Sample prediction over the period February 13, 2015 to December 21, 2023.}
\end{figure}

The LSTM model's performance in predicting volatility was assessed over a time span from  February 13, 2015, to December 21, 2023, for out-of-sample observations. The graph provided in Figure \ref{fig:lstm8_results} illustrates the model's predicted values in comparison with the actual market volatility values. The green solid line represents the LSTM predictions, while the red dashed line denotes the actual values.

The LSTM model's predictions, as visualized in Figure \ref{fig:lstm8_results}, exhibit a high degree of closeness to the actual volatility trend across the observed period. Notably, during the high volatility peak in 2020, the model accurately mirrors the surge in market volatility, underscoring its responsiveness to market dynamics and its ability to capture sudden market movements. However, upon closer examination, particularly during more stable periods such as the span from 2017 to 2018, the model appears to overestimate the volatility. The predictions in these intervals show more fluctuations compared to the actual values, which are relatively smooth. This overestimation suggests that the model may be sensitive to minor variations in the input data, resulting in a higher perceived volatility.

To quantify the performance of the LSTM model, two error metrics were calculated: MAE of \num{1.24e-3} and a RMSE of \num{1.55e-3}. The results are presented in Table \ref{tab:lstm8_metrics}. These low error metrics underscore the model's effectiveness in closely tracking actual market trends, including during periods of significant volatility.

\begin{table}[H]
\centering
\caption{Out-of-Sample Error Metrics for LSTM Model}
\label{tab:lstm8_metrics}
\begin{tabular}{
  >{\raggedright\arraybackslash}p{8cm} 
  S[table-format=1.3e-1] 
  S[table-format=1.3e-1] 
  }
\hline
\textbf{Model} & {\textbf{MAE}} & {\textbf{RMSE}} \\
\hline
LSTM   & 1.24e-3 & 1.55e-3 \\
\hline
\end{tabular}
\caption*{\footnotesize Note: This table presents the LSTM Out of Sample Error Metrics, for walk-forward predictions on a t+1 basis over the period from February 13, 2015, to December 21, 2023.}
\end{table}

\subsection{LSTM-GARCH}
\label{sec:modelling_approaches_lstm_garch}

The hybrid LSTM- GARCH approach enhances the predictive accuracy of volatility forecasting models by extending the LSTM framework to integrate GARCH model predictions as an additional independent variable. This methodology, referred to as hybrid LSTM-GARCH, builds on the foundational LSTM model, which initially incorporates log returns and lagged volatility as predictors. The inclusion of GARCH predictions is based on the hypothesis that the combination of LSTM models' memory capabilities with the volatility modeling strengths of GARCH can yield a more nuanced understanding of market dynamics. 

\subsubsection{LSTM-GARCH Volatility Forecasting Methodology}
\label{sec:volatility_forecasting_methodology_lstm_garch}

The LSTM-GARCH model represents a sophisticated hybrid forecasting approach, specifically designed for financial market volatility. This methodology integrates the strengths of both GARCH and LSTM models to enhance prediction accuracy for future market volatility on the day \(t+1\).

Initially, the GARCH model, detailed in Section \ref{sec:volatility_forecasting-methodology_garch}, is employed to generate preliminary volatility forecasts for day \(t+1\). These forecasts leverage the GARCH model's proficiency in capturing short-term volatility patterns, serving as an initial estimation. Subsequently, these GARCH-generated forecasts are utilized as integral components of the LSTM-GARCH model's input dataset. Along with log returns and lagged volatility values, the GARCH predictions enrich the LSTM-GARCH model's feature set. By incorporating the GARCH model's volatility predictions for day \(t+1\) into the LSTM-GARCH model, this hybrid approach aims to significantly enhance the predictive performance. 

\subsubsection{LSTM- GARCH Results}

\begin{figure}[H]
    \centering
    \caption{LSTM- GARCH Out-of-sample prediction}
    \includegraphics[width=1\linewidth]{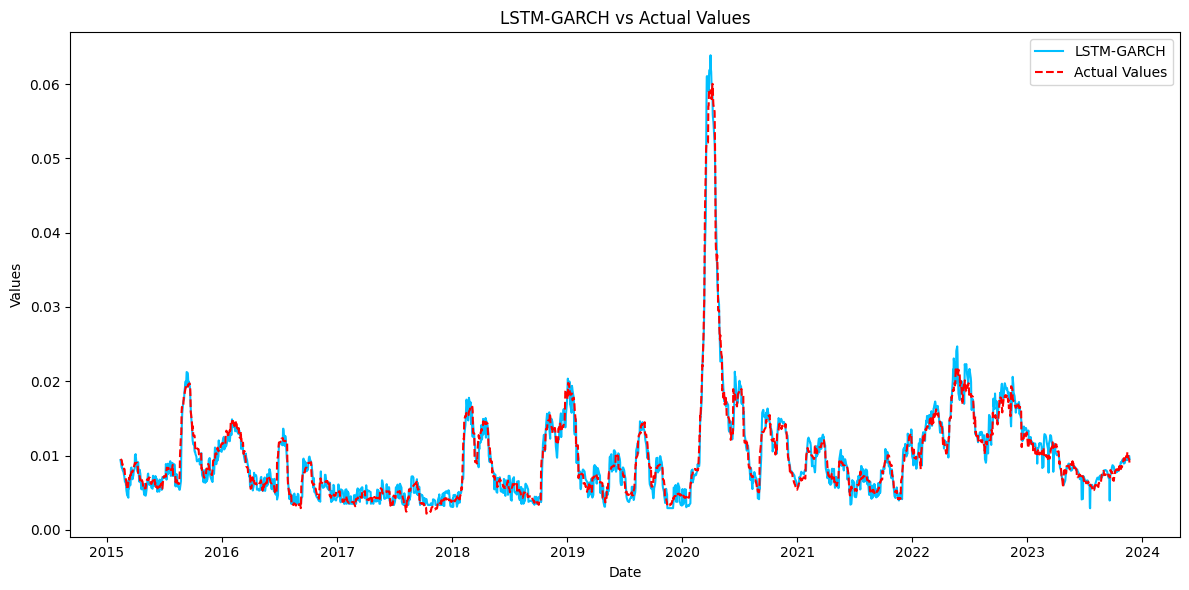}
    \label{fig:lstm9_results}
    \caption*{\footnotesize Note: This figure presents the LSTM- GARCH Out of Sample prediction over the period February 13, 2015 to December 21, 2023.}
\end{figure}

The performance of the LSTM-GARCH model in predicting market volatility was evaluated over the period extending from February 13, 2015, to December 21, 2023. In Figure \ref{fig:lstm9_results}, the model's forecasts are compared with the actual market volatility. The solid line illustrates the predictions generated by the LSTM-GARCH model, while the dashed line represents the actual market volatility.

As observed in Figure \ref{fig:lstm9_results}, the LSTM-GARCH model predictions demonstrate a high correlation with the actual market volatility trajectory throughout the considered timespan. This is especially prominent during the market volatility peak of 2020, wherein the model precisely reflects the sharp increase in volatility, showcasing its acute sensitivity to rapid market shifts and its capability to model drastic changes in market conditions effectively.

Nevertheless, a closer look shows that the LSTM-GARCH model, much like the earlier LSTM model (Section \ref{sec:results_lstm}), sometimes predicts higher volatility than what actually happens. This is especially true for the second half of 2022 and the first half of 2023, where the model's predictions show more ups and downs compared to the actual, more consistent market trends. This similarity suggests that the LSTM-GARCH model, like its LSTM counterpart, might be too responsive to small changes in the data, leading to predictions of higher volatility than is realistic.

To quantify the performance of the LSTM-GARCH model, two error metrics were calculated: MAE of \num{1.01e-3} and an RMSE of \num{1.31e-3}. The results are presented in Table \ref{tab:lstm9_metrics}.

\begin{table}[H]
\centering
\caption{Out-of-Sample Error Metrics for GARCH, LSTM, and LSTM-GARCH Models}
\label{tab:lstm9_metrics}
\begin{tabular}{
    >{\raggedright\arraybackslash}p{8cm}
    S[table-format=1.3e-1]
    S[table-format=1.3e-1]
  }
\hline
\textbf{Model} & {\textbf{MAE}} & {\textbf{RMSE}} \\
\hline
LSTM-GARCH  & 1.01e-3 & 1.31e-3 \\
\hline
\end{tabular}
\caption*{\footnotesize Note: This table presents the LSTM- GARCH Out of Sample Error Metrics, for walk-forward predictions on a t+1 basis over the period from February 13, 2015, to December 21, 2023.}
\end{table}

\subsection{LSTM- GARCH with VIX input}
\label{sec:modelling_approaches_lstm_garch_vix}
Building upon the hybrid LSTM-GARCH framework, the fourth model in this study, referred to as the LSTM-GARCH with VIX input, introduces an additional feature to further refine volatility predictions—the closing prices of the Volatility Index (VIX). The VIX, often referred to as the 'fear index' is a real-time market index representing the market's expectations for volatility over the coming 30 days \citep{CBOE_VIX}. 

The methodological integrity of the LSTM-GARCH approach is preserved in this enhanced model, ensuring that the addition of VIX data complements rather than complicates the forecasting process. By extending the model to include VIX closing prices, the LSTM-GARCH with VIX input model aims to capture not only the quantitative aspects of market data but also the qualitative sentiment reflected by VIX, thus promising a more holistic approach to market volatility forecasting.

\subsubsection{LSTM-GARCH with VIX input Volatility Forecasting Methodology}
\label{sec:volatility_forecasting_methodology_lstm_garch_vix}

Initially, the standard GARCH model processes historical volatility data to produce preliminary forecasts for day \(t+1\). Concurrently, the VIX closing prices (Figure \ref{fig:vix}), alongside log returns (Equation \ref{eq:log_returns}) and lagged volatility (Equation \ref{eq:lagged_volatility}), are fed into the LSTM model, enriching the set of predictive features. The LSTM model then synthesizes the GARCH-derived volatility predictions with the VIX-informed insights to generate a consolidated forecast for a day \(t+1\). 

The introduction of VIX closing prices is expected to enhance the model's capability to capture broader market moods and stress levels, potentially leading to more accurate anticipations of sudden market shifts. Given the VIX's status as a gauge of investor sentiment, its inclusion in the LSTM-GARCH model is hypothesized to offer additional depth to the predictive signals, especially in the face of market uncertainty. 

\subsubsection{LSTM-GARCH with VIX input Results}
\label{sec:lstm10_results}

\begin{figure}[H]
    \centering
    \caption{LSTM- GARCH with VIX input Out-of-sample prediction}
    \includegraphics[width=1\linewidth]{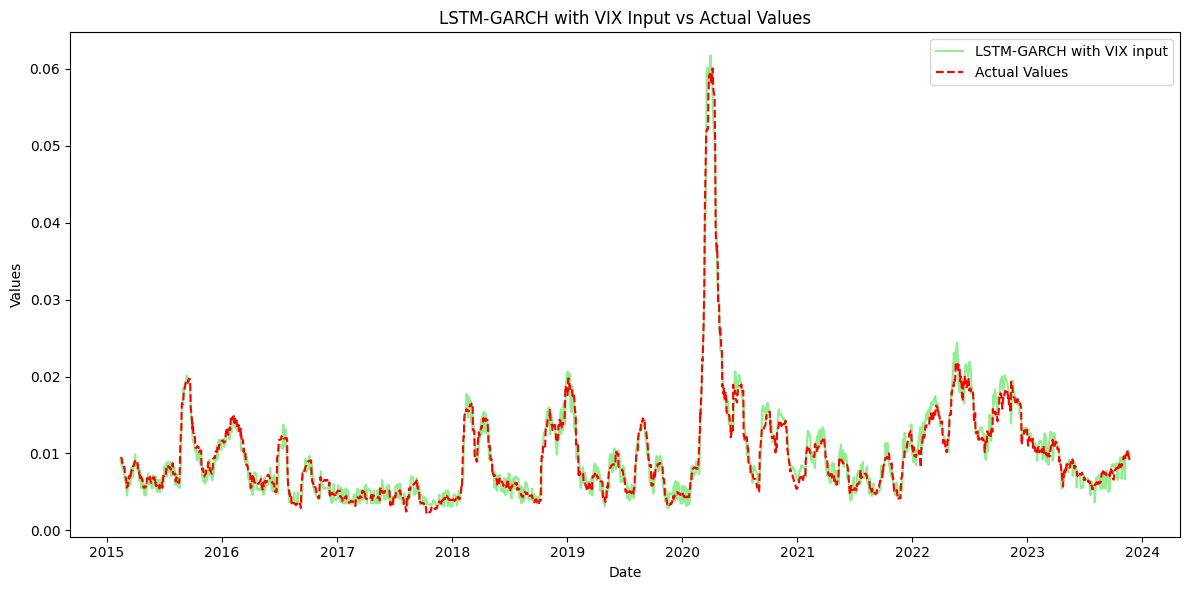}
    \label{fig:lstm10_results}
    \caption*{\footnotesize Note: This figure presents the LSTM- GARCH with VIX input Out of Sample prediction over the period February 13, 2015 to December 21, 2023.}
\end{figure}

Visual inspection of the forecasted versus actual volatility values (Figure~\ref{fig:lstm10_results}) highlights the LSTM-GARCH with the VIX input model's proficiency in capturing volatility spikes, mirroring the significant peaks observed in the 2020 market stress period. The model's forecasts show an elevated alignment with the actual volatility, signifying an improved response to market changes with the addition of VIX data.

Figure~\ref{fig:lstm10_results} also illustrates the model’s ability to smooth out predictions during stable periods, suggesting a mitigation of the overestimation tendency noted in the earlier models. This is indicative of a more balanced sensitivity to input data fluctuations, thanks to the informative nature of VIX pricing.

To quantify the performance of the LSTM-GARCH model with VIX input, two error metrics were calculated: MAE of \num{1.02e-3} and an RMSE of \num{1.30e-3}. The results are presented in Table \ref{tab:lstm10_metrics}.

\begin{table}[H]
\centering
\caption{Out-of-Sample Error Metrics for LSTM-GARCH with VIX Input Models}
\label{tab:lstm10_metrics}
\begin{tabular}{
  >{\raggedright\arraybackslash}p{8cm}
  S[table-format=1.3e-1]
  S[table-format=1.3e-1]
  }
\hline
\textbf{Model} & {\textbf{MAE}} & {\textbf{RMSE}} \\
\hline
LSTM-GARCH with VIX Input   & 1.02e-3  & 1.30e-3 \\
\hline
\end{tabular}
\caption*{\footnotesize Note: This table presents the LSTM- GARCH with VIX input Out of Sample Error Metrics, for walk-forward predictions on a t+1 basis over the period from February 13, 2015, to December 21, 2023.}
\end{table}

\subsubsection{Models comparison}
\label{sec:comparison_results}

The improvements in the error metrics (Table~\ref{tab:error_metrics} and Table~\ref{tab:percentage_improvements}) for the LSTM-GARCH with VIX input model, specifically the lower Root Mean Squared Error (RMSE), suggest a more accurate modeling of market volatility, potentially due to the inclusion of VIX as a sentiment indicator. Compared to the GARCH model, the LSTM-GARCH with VIX input model achieves a lower MAE and RMSE by approximately 34.62\% and 46.03\%, respectively, as the MAE reduces from $1.56 \times 10^{-3}$ to $1.02 \times 10^{-3}$ and the RMSE decreases from $2.39 \times 10^{-3}$ to $1.30 \times 10^{-3}$. Relative to the LSTM model, the LSTM-GARCH with VIX input model records a reduction in RMSE by approximately 16.13\%, with the LSTM model posting an RMSE of $1.55 \times 10^{-3}$. In contrast, there is a slight increase in MAE compared to the LSTM-GARCH model, moving from $1.01 \times 10^{-3}$ to $1.02 \times 10^{-3}$, reflecting a minor degradation of about -0.99\%. This contrast indicates that while the inclusion of VIX improves the RMSE, indicating lower volatility errors, it slightly worsens the average error (MAE) by a small margin. Additionally, the Mann-Whitney U tests (Table~\ref{tab:mann_whitney_results}) confirm the statistical significance of these improvements, highlighting that the differences in MAE are not only numerically but also statistically significant, lending further credence to the superiority of the LSTM-GARCH with VIX input model over the GARCH and LSTM models.

\begin{table}[H]
\centering
\caption{Error Metrics for Selected Models}
\label{tab:error_metrics}
\begin{tabular}{
  >{\raggedright\arraybackslash}p{8cm}
  S[table-format=1.2e-2]
  S[table-format=1.2e-2]
  }
\hline
\textbf{Model} & {\textbf{MAE}} & {\textbf{RMSE}} \\
\hline
GARCH & 1.56e-3 & 2.39e-3 \\
LSTM & 1.24e-3 & 1.55e-3 \\
LSTM-GARCH & 1.01e-3 & 1.31e-3 \\
LSTM-GARCH with VIX input & 1.02e-3 & 1.30e-3 \\
\hline
\end{tabular}
\caption*{\footnotesize Note: This table presents the Out of Sample Error Metrics for our four models: GARCH, LSTM, LSTM-GARCH, and LSTM-GARCH with VIX input. Each t+1 prediction was calculated on a walk-forward basis over the period from February 13, 2015, to December 21, 2023.}
\end{table}

\begin{table}[H]
\centering
\caption{Statistical Comparison of MAE Using Mann-Whitney U Test}
\label{tab:mann_whitney_results}
\begin{tabular}{
  >{\raggedright\arraybackslash}p{10cm}
  S[table-format=7.0] 
  S[table-format=<1.3] 
  }
\hline
\textbf{Model Comparison} & {\textbf{U Statistic}} & {\textbf{p-value}} \\
\hline
LSTM-GARCH with VIX input vs. GARCH & 1963057 & {$<0.001$} \\
LSTM-GARCH with VIX input vs. LSTM & 1991434 & {$<0.001$} \\
LSTM-GARCH with VIX input vs. LSTM-GARCH & 2487891 & 0.257 \\
\hline
\end{tabular}
\caption*{\footnotesize Note: This table presents the results of Mann-Whitney U tests comparing the MAE of the LSTM-GARCH with the VIX input model against other models. A p-value of less than 0.05 indicates a significant difference, suggesting one model has statistically better performance. Significant results highlight the superior accuracy of the LSTM-GARCH with VIX input model over the GARCH and LSTM models.}
\end{table}

\begin{table}[H]
\centering
\caption{Percentage Improvement of LSTM-GARCH with VIX Input Over Other Models}
\label{tab:percentage_improvements}
\begin{tabular}{l S[table-format=2.2] S[table-format=2.2]}
\hline
\textbf{Compared Model} & {\textbf{\% Improvement in MAE}} & {\textbf{\% Improvement in RMSE}} \\
\hline
GARCH & 34.62 & 46.03 \\
LSTM & 17.74 & 16.13 \\
LSTM-GARCH & -0.99 & 0.76 \\
\hline
\end{tabular}
\caption*{\footnotesize Note: This table presents the percentage improvement in MAE and RMSE of the LSTM-GARCH with VIX input model over the other three models: GARCH, LSTM, LSTM-GARCH. Calculations are based on actual improvements recorded between the respective models, reflecting more accurate and updated data.}
\end{table}

Qualitatively, the models differ in their approach to capturing market behavior. The GARCH model, while less complex, may not fully capture the nonlinear patterns present in financial time series data. On the other hand, LSTM networks are known for their ability to learn long-term dependencies, giving them an edge in modeling the temporal aspects of market volatility. The LSTM-GARCH model aims to harness both these strengths, providing a nuanced understanding of market fluctuations. By incorporating VIX as a measure of market sentiment, the LSTM-GARCH with VIX input model extends this capability further, potentially allowing for more informed predictions that consider both historical data and current market sentiment.\\

To deepen our understanding of predictive models' effectiveness across varied market conditions, a systematic analysis was performed by segmenting prediction data based on the volatility present in actual market values. This segmentation, categorizing market conditions into four distinct quartiles—$Lowest$, $Low$-$Medium$, $Medium$-$High$, and $Highest$—enables an examination of model performance across the entire spectrum of market dynamics. These categories were intentionally defined to span the full range of market volatilities, encompassing both periods of relative calm and phases of significant turbulence.

For each volatility quartile, two error metrics, the MAE and the RMSE, were used to evaluate the models' performance (Table \ref{tab:metrics_combined}). By comparing these metrics across the quartiles, as detailed in the accompanying tables, we can assess the predictive accuracy and consistency of each model under different market scenarios. 

For the Lowest Volatility Quartile, the LSTM-GARCH with VIX input model is identified as the best performer with the lowest MAE and RMSE, suggesting its suitability for environments with minimal market fluctuations. Moving to the Low-Medium Volatility Quartile, the LSTM-GARCH model achieves the lowest both MAE and RMSE, advocating its use in slightly more volatile conditions. In the Medium-High Volatility Quartile, the LSTM-GARCH model once again shows the lowest values for both MAE and RMSE, underscoring its robustness in handling medium to high market volatility effectively. Finally, for the Highest Volatility Quartile, this LSTM-GARCH with VIX input model outperforms others in terms of both metrics, reinforcing its suitability for the most volatile market conditions. Overall, the evaluation showcases an ongoing battle between the LSTM-GARCH and the LSTM-GARCH with VIX input models, with each model excelling in different volatility environments. Across the volatility spectrum, the LSTM-GARCH with VIX input model consistently presents as the optimal choice for effectively predicting market movements, especially in low and high-volatility environments. Its balanced performance in terms of both MAE and RMSE across all quartiles makes it the preferred model for general application in varying market conditions.

\begin{table}[H]
\centering
\caption{Error Metrics by Volatility Quartile for All Models}
\label{tab:metrics_combined}
\small 
\begin{tabular}{
  >{\raggedright\arraybackslash}p{10cm}
  S[table-format=1.3e-1, detect-weight]
  S[table-format=1.3e-1, detect-weight]
  }
\hline
\textbf{Model} & {\textbf{MAE}} & {\textbf{RMSE}} \\
\hline
\multicolumn{3}{c}{\textbf{Lowest Volatility Quartile}} \\
\hline
GARCH                        & 1.42e-3 & 1.53e-3 \\
LSTM                         & 1.06e-3 & 1.19e-3 \\
LSTM-GARCH                   &  7.95e-4 &  8.97e-4 \\
LSTM-GARCH with VIX Input    &\bfseries 7.63e-4 & \bfseries 8.73e-4 \\
\hline
\multicolumn{3}{c}{\textbf{Low-Medium Volatility Quartile}} \\
\hline
GARCH                        & 8.96e-4 & 1.17e-3 \\
LSTM                         & 1.11e-3 & 1.30e-3 \\
LSTM-GARCH                   & \bfseries 8.62e-4 & \bfseries 1.05e-3 \\
LSTM-GARCH with VIX Input    & 9.19e-4 & 1.07e-3 \\
\hline
\multicolumn{3}{c}{\textbf{Medium-High Volatility Quartile}} \\
\hline
GARCH                        & 1.20e-3 & 1.54e-3 \\
LSTM                         & 1.26e-3 & 1.53e-3 \\
LSTM-GARCH                   & \bfseries 1.05e-3 &  \bfseries 1.33e-3 \\
LSTM-GARCH with VIX Input    & 1.08e-3 &  1.35e-3 \\
\hline
\multicolumn{3}{c}{\textbf{High Volatility Quartile}} \\
\hline
GARCH                        & 2.69e-3 & 4.07e-3 \\
LSTM                         & 1.40e-3 & 1.95e-3 \\
LSTM-GARCH                   & 1.32e-3 & 1.79e-3 \\
LSTM-GARCH with VIX Input    & \bfseries 1.30e-3 & \bfseries 1.73e-3 \\
\hline
\end{tabular}
\caption*{\footnotesize Note: This table presents Out of Sample Error Metrics by Volatility Quartile. The best values for MAE and RMSE in each quartile are highlighted in bold. The Out of Sample observations were categorized based on volatility magnitude and they were divided into four distinct quartiles: Lowest, Low-Medium, Medium-High, and High.}
\end{table}

Additionally, the study employs a directional accuracy test across the four models, evaluating their ability to predict market trend directions over periods of 1, 5, and 22 days. This assessment quantifies each model's precision in forecasting the trend's course, thus providing a detailed view of their predictive power in short-, medium-, and longer-term market movements. It is crucial to note that these models are fundamentally designed for regression tasks rather than the classification of volatility direction. This distinction highlights their intent to predict continuous outcomes, providing a deeper analytical framework for understanding market behaviors rather than merely categorizing trend directions. Thus, the results from Table \ref{tab:directional_accuracy} are for informational purposes.

Short-term (1-day) forecasts show higher accuracy for the LSTM and LSTM-GARCH with VIX input models, suggesting their effectiveness in capturing immediate market trends. The accuracy generally increases for the medium-term (5-day) forecasts, except for the LSTM model which performs slightly worse, indicating their robustness over slightly longer durations. For long-term (22-day) forecasts, the accuracies tend to converge, with no model showing a distinct advantage, highlighting the inherent complexities in long-duration market prediction, where LSTM's performance notably declines, reflecting the escalating difficulty in sustaining prediction accuracy over extended periods.

\begin{table}[H]
\centering
\caption{Directional Prediction Accuracy of Models for Different Periods}
\label{tab:directional_accuracy}
\begin{tabular}{
  >{\raggedright\arraybackslash}p{6cm}
  S[table-format=2.2, detect-weight]
  S[table-format=2.2, detect-weight]
  S[table-format=2.2, detect-weight]
  }
\hline
\textbf{Model} & {\textbf{1-Day (\%)}} & {\textbf{5-Day (\%)}} & {\textbf{1-Month (22 Days) (\%)}} \\
\hline
GARCH & 50.73 & 53.58 & 51.31 \\
LSTM & \bfseries 55.31 & 55.12 & 49.39 \\
LSTM-GARCH & 54.97 & 57.72 &  50.77 \\
LSTM-GARCH with VIX Input & 55.02 & \bfseries 59.27 &  \bfseries 51.94 \\
\hline
\end{tabular}
\caption*{\footnotesize Note: This table presents the directional prediction accuracy for 1, 5, and 22 days ahead. It is crucial to note that these models are fundamentally designed for regression tasks
rather than the classification of volatility direction, hence these results are only for informational purposes.}
\end{table}

\section{Sensitivity Analysis}
\label{sensitivity}

The sensitivity of the LSTM-GARCH model with VIX input to its setup and parameters was closely examined. This analysis specifically aimed to understand the impact of alterations in the loss function, sequence length, number of layers in the LSTM model, and the type of activation functions. In each scenario analyzed, a single parameter was modified with the remainder of the model's configuration held constant to isolate the effects of each change.  The below scenarios were considered: 

\begin{itemize}
    \item Use MAE as a loss function instead of MSE
    \item Replace input Log Returns with Daily Percentage Changes
    \item Sequence length
    \begin{itemize}[label={--}]
        \item Decrease sequence length to 5 days
        \item Increase sequence length to 66 days
    \end{itemize}
    \item LSTM Architecture Number of Layers
    \begin{itemize}[label={--}]
        \item Reduce the number of layers to 1
        \item Increase the number of layers to 3
    \end{itemize}
\end{itemize}
\begin{itemize}
\item ReLu Activation Function for both LSTM Layers

\end{itemize}

\begin{table}[H]
\centering
\caption{Error Metrics for LSTM-GARCH with VIX Input Model Under Different Sensitivity Cases}
\label{tab:metrics_sensitivity}
\begin{tabular}{
  >{\raggedright\arraybackslash}p{10cm}
  S[table-format=1.3e-1, detect-weight]
  S[table-format=1.3e-1, detect-weight]
  }
\hline
\textbf{Sensitivity Case} & {\textbf{MAE}} & {\textbf{RMSE}} \\
\hline
Use MSE as a loss function * & 1.02e-3 & 1.30e-3 \\
Use MAE as a loss function & 1.02e-3 & 1.33e-3 \\
\hline
Log Returns as Input * & 1.02e-3 & 1.30e-3 \\
Percentage Change in Price as Input & \bfseries 9.80e-4 & \bfseries 1.26e-3 \\
\hline
Sequence length of 22 days * & 1.02e-3 & 1.30e-3 \\
Decrease sequence length to 5 days & \bfseries 8.90e-04 & \bfseries 1.19e-3 \\
Increase sequence length to 66 days & 1.09e-3 & 1.38e-3 \\
\hline
Dual- Layer LSTM Architecture *& 1.02e-3 & 1.30e-3 \\
Single- Layer LSTM Architecture & \bfseries 8.23e-4 & \bfseries 1.15e-3 \\
Three- Layer LSTM Architecture & 1.23e-3 & 1.50e-3 \\
\hline
Activation function: 1st Layer- Tanh, 2nd Layer- Tanh * & 1.02e-3 & 1.30e-3 \\
Activation function: 1st Layer- ReLU, 2nd Layer- ReLU & 1.16e-3 & 1.60e-3 \\
\hline
\end{tabular}
\caption*{\footnotesize Note: This table presents the Out of Sample Error Metrics for the chosen sensitivity cases for our best performing LSTM- GARCH with VIX input model. The (*) denotes the setup that was used for the base model. The bolded font indicates values better than for the base case scenario.}
\end{table}

\subsection{Sensitivity- Using MAE as a loss function instead of MSE}

\begin{figure}[H]
    \centering
    \caption{LSTM-GARCH with VIX Input sensitivity to loss function}
    \includegraphics[width=1\linewidth]{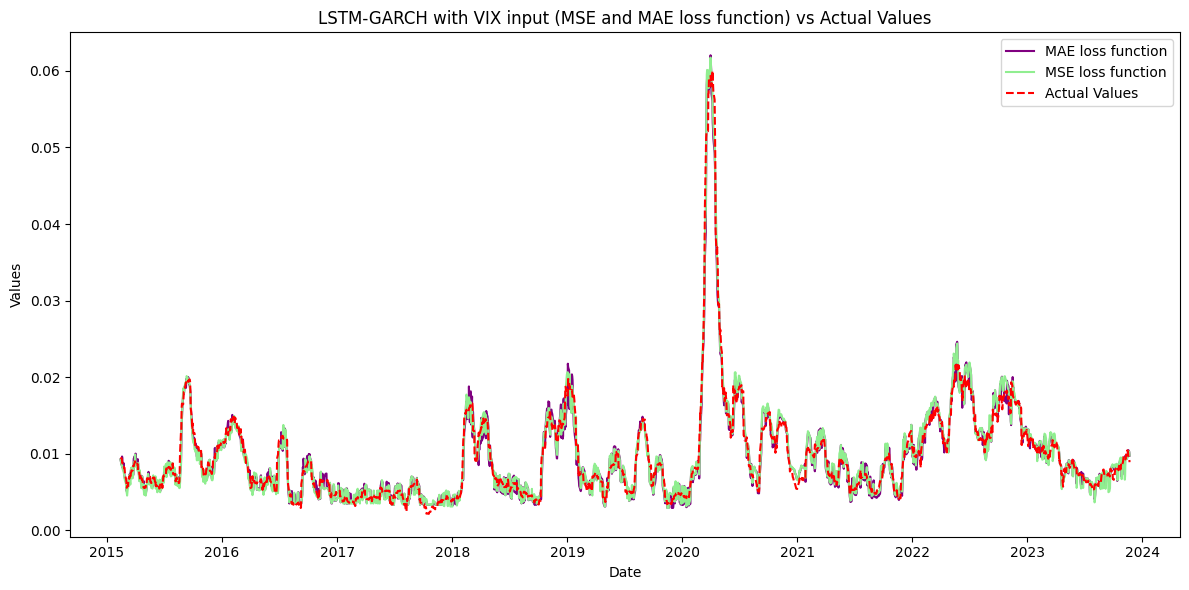}
    \label{fig:sensitivity_loss_func}
    \caption*{\footnotesize Note: This figure presents the sensitivity of LSTM- GARCH with VIX input model to different loss functions: MAE (blue) and MSE (green) over the period February 13, 2015 to December 21, 2023.}
\end{figure}

The sensitivity analysis of the LSTM-GARCH with VIX input model with respect to the choice of the loss function is illustrated in Table~\ref{tab:metrics_sensitivity} and Figure~\ref{fig:sensitivity_loss_func}. When transitioning from the MSE to MAE as the loss function, a slight increase in both the MAE and RMSE is observed, suggesting that the model is more precise with MSE. This could be attributed to MSE's propensity to heavily penalize larger errors, thereby being more sensitive to outliers than MAE \citep{chai2014}. The subtleties of this effect are captured in the graphical representation (Figure~\ref{fig:sensitivity_loss_func}), where the predictions under MSE adherence exhibit a closer adherence to actual values. The findings underscore the significance of aligning the loss function with the underlying error distribution and the specific requirements of the predictive modeling task at hand.

\subsection{Sensitivity- Replacing Log Returns with Daily Percentage Changes}

\begin{figure}[H]
    \centering
    \caption{LSTM- GARCH with VIX input sensitivity to different forms of input data: Log Returns versus simple Percentage Changes in Price}
    \includegraphics[width=1\linewidth]{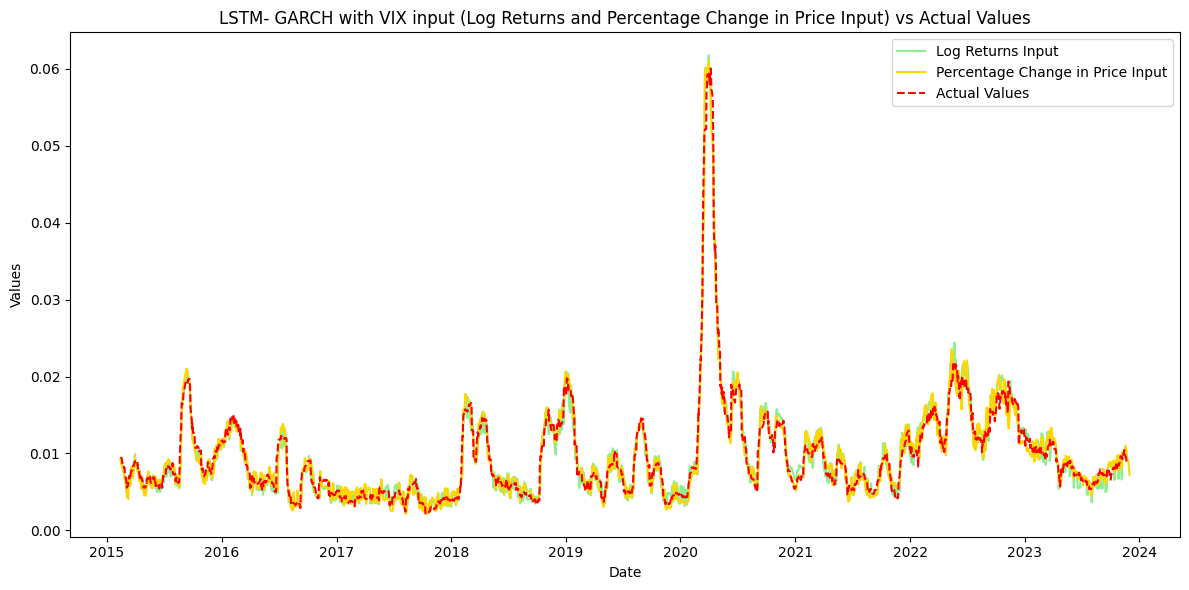}
    \label{fig:sensitivity_log_perctentage}
    \caption*{\footnotesize Note: This figure presents the sensitivity of LSTM- GARCH with VIX input model to different forms of input data: percentage change in price (green) and log returns (yellow) over the period February 13, 2015 to December 21, 2023.}
\end{figure}

The LSTM-GARCH model's performance under various input data types was scrutinized, with a focus on log returns and percentage changes in price. As illustrated in Figure~\ref{fig:sensitivity_log_perctentage}, the model exhibited notable consistency when leveraging log returns as input data. In contrast, the use of percentage changes in price as input resulted in marginally lower MAE and RMSE, as can be seen in  Table~\ref{tab:error_metrics}. This marginal reduction indicates a potential preference for percentage changes in price over log returns when precision is paramount. However, this preference is model and context-specific, considering log returns generally provide a more stable and normalized measure for financial time series data \citep{tsay2010analysis}. 

\subsection{Sensitivity- Sequence Length} 

\begin{figure}[H]
    \centering
    \caption{LSTM-GARCH with VIX Input sensitivity to sequence length}
    \includegraphics[width=1\linewidth]{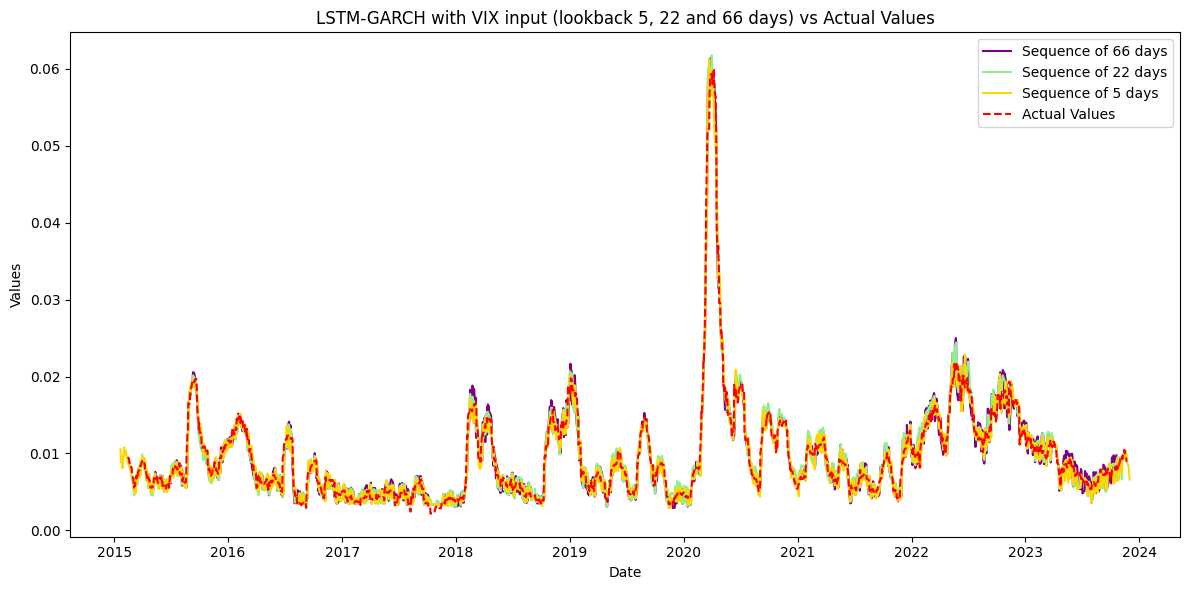}
    \label{fig:sensitivity_sequence_length}
    \caption*{\footnotesize Note: This figure presents the sensitivity of LSTM- GARCH with VIX input model to sequence length: 66 days (green), 5 days (yellow), 22 days (blue) over the period February 13, 2015 to December 21, 2023.}
\end{figure}

A detailed assessment was carried out to examine how well the LSTM-GARCH with VIX input model performs with different input sequence lengths, specifically comparing 5-day and 66-day periods against the chosen 22-day period. The findings, presented in Table~\ref{tab:metrics_sensitivity} and Figure \ref{fig:sensitivity_sequence_length}, showed that the model consistently provided accurate predictions across these variations, indicating its flexibility. 

Shortening the sequence to 5 days decreased the MAE and RMSE, implying a rapid reaction to market volatility. Conversely, an extended 66-day sequence saw a marked increase in errors, indicating a potential dilution of predictive relevancy with overextended historical data incorporation. This analysis underscores the importance of selecting an appropriate sequence length that reflects the desired balance between responsiveness to recent market events and the incorporation of long-term trends, an aspect further discussed in the survey by \citep{dama2021timeseries}.

\subsection{Sensitivity- Number of LSTM layers}

We also studied how the architecture's number of LSTM layers influences the model's predictive accuracy. The foundational structure of the base model includes two LSTM layers, each with 128 neurons and a 'tanh' activation function, interspersed with dropout layers set at a rate of 0.1 to mitigate overfitting. This dual-layer configuration is the benchmark against which other architectural modifications were measured.

\begin{figure}[H]
    \centering
    \caption{LSTM- GARCH with VIX input sensitivity to the number of LSTM layers}
    \includegraphics[width=1\linewidth]{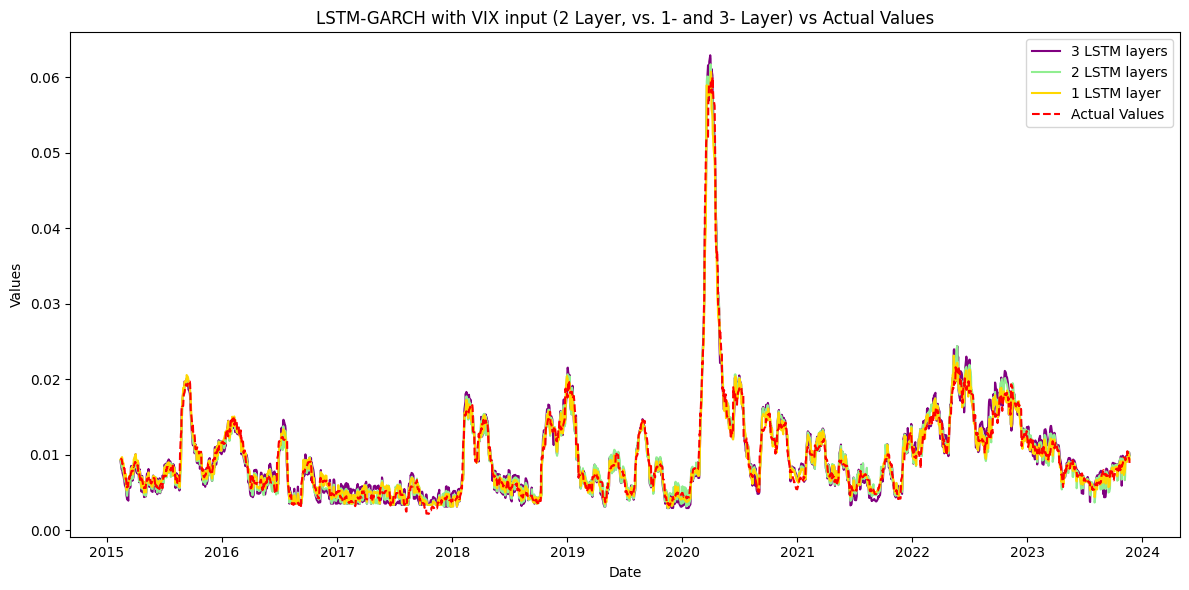}
    \label{fig:sensitivity_layers}
    \caption*{\footnotesize Note: This figure presents the sensitivity of LSTM- GARCH with VIX input model to the number of LSTM hidden layers: 3 layers (purple), 1 layer (green), 2 layers (yellow) over the period February 13, 2015 to December 21, 2023.}
\end{figure}

Results from the sensitivity analysis, as shown in Table~\ref{tab:metrics_sensitivity} and Figure~\ref{fig:sensitivity_layers}, reveal that reducing the model to a single-layer LSTM architecture not only improves computational efficiency but also enhances performance, with a decrease in MAE to \(8.23 \times 10^{-4}\) and RMSE to \(1.15 \times 10^{-3}\). This suggests that one LSTM layer is sufficient for capturing the essential temporal dependencies in the data, indicating that additional layers might lead to overfitting or unnecessary complexity without proportionate gains in accuracy. In contrast, expanding the model to include a third LSTM layer results in elevated error metrics, with the MAE increasing to \(1.23 \times 10^{-3}\) and RMSE to \(1.50 \times 10^{-3}\). This heightened error suggests that the additional layer could be introducing unnecessary complexity, potentially leading to overfitting and a diminished ability to generalize from the input data. The sensitivity analysis thus highlights the effectiveness of the single-layer LSTM model in outperforming the multi-layer configurations by providing a more parsimonious, yet robust approach to modeling financial time series data.

\subsection{Sensitivity- Activation Functions}

\begin{figure}[H]
    \centering
    \caption{LSTM-GARCH with VIX Input sensitivity to the choice of activation function}
    \includegraphics[width=1\linewidth]{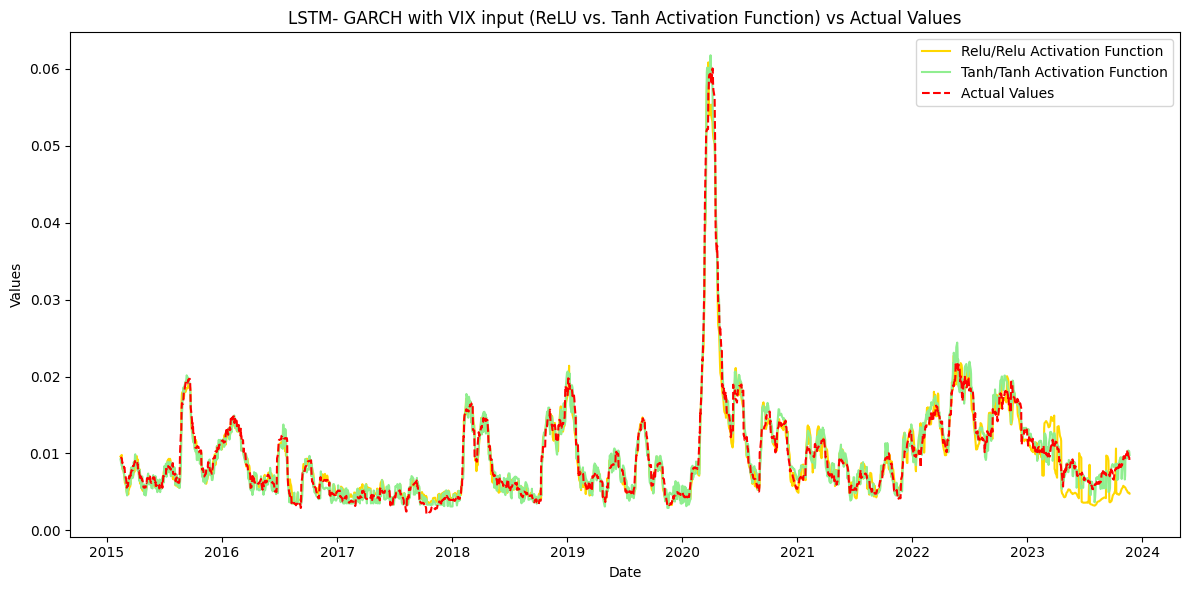}
    \label{fig:sensitivity_activation_function}
    \caption*{\footnotesize Note: This figure presents the sensitivity of LSTM- GARCH with VIX input model to the choice of activation function for the two hidden layers: ReLU/ Tanh (blue), ReLU/ ReLU (green), Tanh/ Tanh (yellow) over the period February 13, 2015 to December 21, 2023.}
\end{figure}

In analyzing the sensitivity of the LSTM-GARCH model to various activation functions, a distinct impact on performance metrics is observed in Table~\ref{tab:metrics_sensitivity} and Figure~\ref{fig:sensitivity_activation_function}. The standard configuration utilizing hyperbolic tangent (tanh) activation functions for both LSTM layers results in a MAE of \(1.02 \times 10^{-3}\) and an RMSE of \(1.30 \times 10^{-3}\), which serves as a baseline for comparison. Modifying both layers to a Rectified Linear Unit (ReLU) activation function leads to an increased MAE of \(1.16 \times 10^{-3}\) and RMSE of \(1.60 \times 10^{-3}\). This reduction in prediction accuracy may be attributable to the unbounded nature of ReLU, which can potentially result in gradient instability.

The tanh nonlinearity is particularly advantageous because its outputs are zero-centered, contributing to more predictable and stable gradient descent trajectories during training. Unlike tanh, the ReLU function can cause gradients to either explode or vanish during backpropagation, particularly when dealing with deep networks or complex patterns. Moreover, the zero-centered nature of tanh outputs tends to yield more efficient learning progress in layers deep within a network. This characteristic of tanh is beneficial for modeling complex nonlinear relationships, such as those found in financial time series data, enhancing the reliability of the predictions without the risks associated with the unbounded nature of ReLU \citep{goodfellow2016deep}.

\section{LSTM Explainability through LIME}

Local Interpretable Model-agnostic Explanations (LIME) revolutionize the interpretability of machine learning models, making the inner workings of complex models like Long Short-Term Memory (LSTM) networks more transparent. LIME, as introduced by \citep{Ribeiro2016}, approximates complex model predictions locally with interpretable models by generating and analyzing permutations of input data. This approach, particularly beneficial for 'black box' models, illuminates the influence of individual features on predictions, enhancing our understanding of model decisions. LIME is characterized by three foundational principles:

\begin{enumerate}
    \item \textbf{Interpretable}: It generates explanations that are easy for humans to understand, using simple models to approximate the predictions of potentially complex machine learning models.
    \item \textbf{Local Fidelity}: LIME focuses on providing accurate explanations in the local context around the prediction, ensuring that the explanations are faithful to what the model computes in the vicinity of the instance being examined.
    \item \textbf{Model Agnostic}: Its application is not limited by the type of machine learning model, making LIME versatile and widely applicable across various models and algorithms.
\end{enumerate}

LIME generates local surrogate models that closely mimic a complex model's behavior within a specified vicinity, thereby revealing the contribution of each feature to the prediction. This is achieved through data permutation and observing prediction changes, which are then used to train a simpler model, such as Lasso or a decision tree. The mathematical essence of LIME, presented in Equation \ref{eq:lime}, seeks to minimize the loss function $L$, representing the difference between the complex model $f$ and the interpretable model $g$, while maintaining the simplicity of the explanation model through $\Omega(g)$.

\begin{equation}
 \label{eq:lime}
\text{explanation}(x) = \arg\min_{g \in G} L(f, g, \pi_x) + \Omega(g)
\end{equation}

Here, $g$ is the interpretable model chosen to minimize $L$, such as mean squared error, which measures the approximation accuracy of the explanation to $f$'s prediction. The complexity $\Omega(g)$ is kept low to prefer simpler explanations. The set $G$ encompasses all potential explanatory models, with $\pi_x$ defining the proximity of sampled instances to the instance of interest, $x$. LIME's optimization primarily addresses the loss, allowing users to adjust model complexity, for example, by limiting the number of features. 

In applying LIME to our LSTM- GARCH with VIX input model for explainability, we first preprocessed our dataset for compatibility, scaling features, and reshaping data to fit LSTM's input structure. For LIME's analysis, we adapted our LSTM's input by flattening the 3D time-sequenced data into a 2D format. Utilizing the LimeTabularExplainer, we crafted a custom prediction function that reshapes and scales input data accordingly, enabling LIME to probe our LSTM- GARCH with VIX input model effectively. 

\begin{figure}[H]
    \centering
    \caption{LIME local explainability for February 13, 2015}
    \includegraphics[width=1\linewidth]{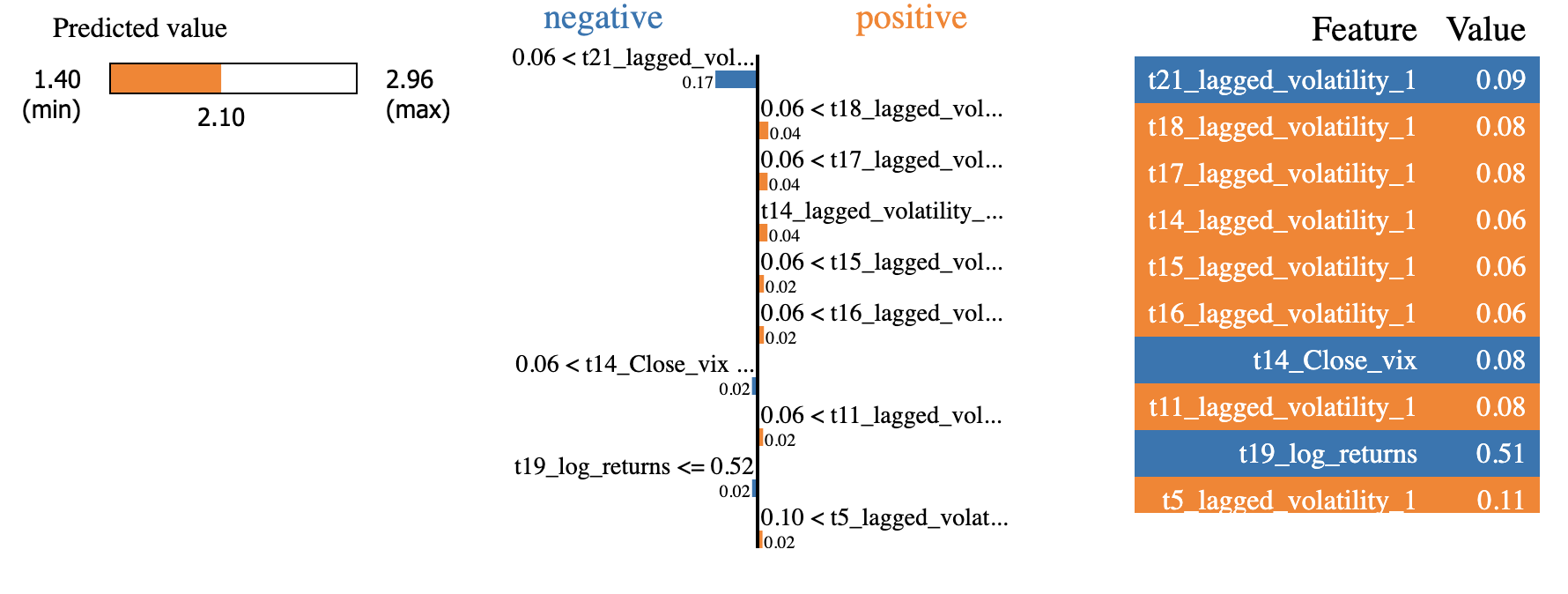}
    \label{fig:lime_2015_02_13}
    \caption*{\footnotesize Note: This figure presents the LIME local explainability for February 13, 2015. On the left, the 'Predicted value' section shows the predicted median volatility of 2.10, with a range from 1.40 to 2.96. In the middle, the 'Negative and Positive' section highlights the features that either decrease or increase the predicted volatility. On the right, the 'Feature Value' section lists the actual values of these influential features on the specific date.
}
\end{figure}

As a use case example, the visualization provided by LIME for February 13, 2015, (Figure~\ref{fig:lime_2015_02_13}) illustrates the impact of various features on the LSTM model's prediction of market volatility. The 'Predicted value' section (located on the left side in Figure~\ref{fig:lime_2015_02_13}) displays the predicted median value of volatility for the analyzed date, which is 2.10, with the prediction interval ranging from a minimum of 1.40 to a maximum of 2.96. The 'Negative and Positive' section (situated in the middle of Figure~\ref{fig:lime_2015_02_13}) identifies the features contributing to a decrease (left - negative) or increase (right - positive) in the predicted value, along with their respective weights. These features have been identified as significantly influential for this specific local prediction on the selected day. In detail, features such as $0.06 < \text{t21 lagged\_volatility}$, $0.06 < \text{t14 Close\_vix}$, and $\text{t19 log\_returns} \leq 0.52$ are shown to negatively impact the volatility prediction, indicating an association with lower volatility levels. Conversely, $0.06 < \text{t18 lagged\_volatility}$ has a positive impact on the predicted volatility, suggesting a link with higher volatility. The 'Feature Value' section presents the actual values of the top features for the specific date. For instance, the value of $\text{t21 lagged\_volatility}$ is 0.09, which is greater than 0.06, thereby negatively influencing the predicted value as per the model's analysis.

Despite the invaluable perspectives LIME provides on model interpretability, it is not without its limitations. The balance between simplicity and accuracy in the interpretable models used poses a significant challenge, as these models may not capture the full complexity of the original predictions. This simplification can lead to partially accurate or oversimplified explanations. Moreover, the local nature of LIME's explanations limits their applicability to a broader data set or differing contexts, constraining the generalizability of the insights offered. The assumptions underlying feature selection and perturbation generation also introduce potential biases, further highlighting the importance of critical evaluation and understanding of LIME-generated explanations within specific application contexts. \\

\section{Conclusion}
This paper provided a comprehensive examination of volatility forecasting for the S\&P 500 index through the application and integration of various models, including GARCH, LSTM, hybrid LSTM-GARCH, and an advanced hybrid model incorporating VIX inputs. Each model was designed to leverage different aspects of the data, from historical volatility patterns captured by GARCH models to deep learning's capability to recognize complex nonlinear patterns in the LSTM and hybrid models. The inclusion of the VIX index in the final hybrid model aimed to introduce market sentiment into the forecasting process, thereby enriching the models' contextual awareness and predictive accuracy.

Our findings revealed that hybrid models, especially those incorporating VIX inputs, significantly outperformed traditional GARCH models. These advanced models demonstrated superior capability in managing the complexities inherent in financial time series data, showcasing enhanced predictive accuracy and robustness. The integration of LSTM networks with GARCH models allowed for an effective synthesis of short-term historical volatility and long-term dependencies, which was further augmented by the real-time sentiment analysis provided by the VIX index. \\

The following conclusions are drawn from the hypothesis testing conducted in this study:

\begin{itemize}
    \item \textbf{RH1:} The results support the hypothesis (RH1) that market prices do not fully reflect all available information, indicating some level of predictability in market movements.
    
    \item \textbf{RH2:} The GARCH model was found to effectively identify patterns of historical volatility, supporting the hypothesis (RH2) and establishing it as a reliable benchmark for newer forecasting approaches.
    
    \item \textbf{RH3:} LSTM networks were confirmed to outperform traditional models like GARCH in forecasting the S\&P 500 index's volatility, supporting the hypothesis (RH3).
    
    \item \textbf{RH4:} The hybrid LSTM-GARCH model surpassed the performance of standalone LSTM and GARCH models, validating the hypothesis (RH4).
    
    \item \textbf{RH5:} Including VIX inputs enhanced the accuracy of volatility forecasts, supporting the hypothesis (RH5).
    
    \item \textbf{RH6:} The use of the Local Interpretable Model-agnostic Explanations (LIME) technique enhanced the interpretability of the hybrid LSTM-GARCH model with VIX input, substantiating the hypothesis (RH6).
\end{itemize}

\textbf{Results from Side Hypotheses for Sensitivity Testing:}
\begin{itemize}
    \item \textbf{sRH1:} Changing the loss function from MSE to MAE did not lead to more robust forecasts, refuting this specific hypothesis.
    \item \textbf{sRH2:} Replacing input Log Returns with Daily Percentage Changes improved the accuracy of volatility forecasts, supporting this hypothesis.
    \item \textbf{sRH3:} Decreasing the sequence length to 5 days had mixed results: it improved the MAE but slightly worsened the RMSE, providing partial support for the hypothesis.
    \item \textbf{sRH4:} Increasing the sequence length to 66 days did not improve the accuracy of forecasts, refuting this hypothesis.
    \item \textbf{sRH5:} Reducing the number of LSTM layers to 1 had mixed outcomes: it improved the MAE but worsened the RMSE, leading to partial support for the hypothesis.
    \item \textbf{sRH6:} Increasing the number of LSTM layers to 3 worsened the model’s performance significantly, contradicting the hypothesis.
    \item \textbf{sRH7:} The performance of LSTM models was significantly impacted by the choice of activation function, confirming this hypothesis.
\end{itemize}

This study introduced significant advancements in financial volatility forecasting by pioneering a hybrid approach that combines LSTM and GARCH models. This integration enhances prediction accuracy by incorporating the VIX index to reflect market sentiment. Additionally, we applied the Local Interpretable Model-agnostic Explanations (LIME) technique to the LSTM- GARCH with VIX Input model, offering insights into its decision-making processes and improving the interpretability of deep learning in finance. These contributions provide valuable frameworks for both academic research and practical applications in financial risk management.

Future research could expand in several promising directions to build on the groundwork laid by this paper. Firstly, incorporating macroeconomic indicators such as GDP growth rates, unemployment rates, and inflation figures could provide a more holistic view of the factors influencing market volatility. These macroeconomic factors could be integrated into the existing models to examine their predictive power in conjunction with market data. Secondly, the exploration of advanced machine learning architectures such as Transformer models, which have excelled in processing sequential data in other domains, could offer new insights and methodologies for financial time series analysis. These models could be particularly adept at handling the multi-scale temporal dynamics often observed in market data. Thirdly, enhancing the models to process real-time data would allow for dynamic updating of predictions, a critical feature for applications in high-frequency trading and real-time risk management. This would involve developing techniques for online learning where the model continuously updates itself as new data becomes available. Furthermore, extending the hybrid modeling approach to other financial indices or asset classes could provide a comparative analysis of model performance across different markets. This would help in understanding if the models' effectiveness is consistent across various market conditions and asset types. Lastly, deepening the focus on model interpretability and explainability is crucial, especially in finance where decision-making involves significant economic stakes. Advanced techniques in explainable artificial intelligence could be applied to these models to uncover the underlying decision processes, thereby making the models more transparent and trustworthy for users. These areas not only promise to refine the predictive capabilities of volatility models but also enhance their applicability in practical financial settings, supporting more informed and effective decision-making in the finance industry.

\bibliographystyle{plainnat}
\bibliography{bibliography}

\end{document}